\documentclass[epj]{svjour}

\usepackage{dsfont}
\usepackage{graphicx}
\usepackage{epstopdf}
\usepackage{epsfig}
\usepackage{amsfonts}
\usepackage{amssymb, graphics, amsmath}
\usepackage{hyperref}
\usepackage{color}
\usepackage{pdflscape}
\usepackage{pifont}
\usepackage{bm}
\usepackage{slashed}

\usepackage{stfloats}
\usepackage{widetext}
\newcommand*{\rom}[1]{\expandafter\romannumeral #1}
\newcommand*{\Rom}[1]{\uppercase\expandafter{\romannumeral #1\relax}}

\begin{document}
\title{The electromagnetic Sigma-to-Lambda transition form factors with coupled-channel
  effects in the space-like region}

\author{Yong-Hui Lin\inst{1}
\and Hans-Werner Hammer\inst{2,3}
\and Ulf-G.~Mei{\ss}ner\inst{1,4,5}
}

\institute{
Helmholtz-Institut~f\"{u}r~Strahlen-~und~Kernphysik~and~Bethe~Center~for
Theoretical~Physics, Universit\"{a}t~Bonn, \\ D-53115~Bonn,~Germany
\and
Technische Universit\"at Darmstadt, Department of Physics,
Institut f\"ur Kernphysik,\\ 64289 Darmstadt, Germany
\and
ExtreMe Matter Institute EMMI and Helmholtz Forschungsakademie Hessen f\"ur
FAIR (HFHF),\\ GSI Helmholtzzentrum f\"ur Schwerionenforschung GmbH,
64291 Darmstadt, Germany
\and
Institut~f\"{u}r~Kernphysik,~Institute~for~Advanced~Simulation and
J\"{u}lich~Center~for~Hadron~Physics, \\ 
Forschungszentrum~J\"{u}lich, D-52425~J\"{u}lich,~Germany
\and Tbilisi State University, 0186 Tbilisi, Georgia}

\titlerunning{Sigma-to-Lambda transition form factors}
\authorrunning{Lin, Hammer, Mei{\ss}ner}
\date{Received: date / Revised version: date}
%

\abstract{Using dispersion theory, the electromagnetic Sigma-to-Lambda transition form factors
are expressed as the product of the pion electromagnetic form factor and the
$\Sigma\bar{\Lambda}\to\pi\pi$
scattering amplitudes with the latter estimated from SU(3) chiral perturbation theory including the
baryon decuplet as explicit degrees of freedom. The contribution of the $K\bar{K}$ channel is also
taken into account and the $\pi\pi$-$K\bar{K}$ coupled-channel effect is included by means of
a two-channel Muskhelishvili-Omn\`{e}s representation. It is found that the electric transition
form factor shows a significant shift after the inclusion of the $K\bar{K}$ channel, while the
magnetic transition form factor is only weakly affected. However, the $K\bar{K}$ effect on the electric
form factor is obscured by the undetermined coupling $h_A$ in the three-flavor chiral Lagrangian. The error bands
of the Sigma-to-Lambda transition form factors from the uncertainties of the couplings and low-energy constant in 
three-flavor chiral perturbation theory are estimated by a bootstrap sampling method. 
}

\maketitle

\section{Introduction}\label{sec:intro}

Electromagnetic form factors (EMFFs) give access to the  strong
interaction, which provides one of the most notorious challenges in the Standard Model due to the
nonperturbative nature of Quantum Chromodynamics~(QCD) at the low energy scale. On the one hand,
the EMFFs can be extracted from a variety of experimental processes, such as  lepton-hadron
scattering,  lepton-antilepton annihilation or radiative hadron decays. These EMFFs can be measured
over a large  energy range. On the other hand, dispersion theory, which is a powerful nonperturbative approach,
allows for a theoretical description of the EMFFs. Consequently, the EMFFs are an ideal bridge
between experimental measurements and theoretical studies of the low-energy strong interaction.

In the last decade, much research effort both in experiment and theory was focused
on the nucleon EMFFs, largely triggered by the so-called proton radius puzzle~\cite{Pohl:2010zza}.
For recent reviews, see e.g. Refs.~\cite{Denig:2012by,Pacetti:2014jai,Punjabi:2015bba,Peset:2021iul,Gao:2021sml}.
In the process of unravelling this puzzle,  dispersion theory has  played and is playing  a crucial role in the
theoretical description of the nucleon
EMFFs~\cite{Lorenz:2012tm,Lorenz:2014yda,Hoferichter:2016duk,Lin:2021umk}. The dispersion theoretical
parametrization of the nucleon EMFFs, first proposed in the early works~\cite{Chew:1958zjr,Federbush:1958zz,Hohler:1976ax} and further developed in Refs.~\cite{Mergell:1995bf,Belushkin:2006qa,Hoferichter:2016duk},
incorporates all constraints from unitarity, analyticity, and crossing symmetry, as well as the
constraints on the asymptotic behavior of the form factors from perturbative QCD~\cite{Lepage:1980fj}.
The state of the art of dispersive analyses of the nucleon EMFFs is reviewed in
Ref.~\cite{Lin:2021umz}. Very recently, all current measurements on  electron-proton scattering,
electron-positron annihilation,  muonic hydrogen spectroscopy, and polarization measurements from
Jefferson Laboratory could be consistently described  in a dispersion theoretical analysis of
the nucleon EMFFs~\cite{Lin:2021xrc}. 

The dispersive prescription of parameterizing the nucleon EMFFs can also be applied to
other hadron states. The first two straightforward extensions concern  the Delta baryon and
the hyperon states, with the former obtained by flipping the spin of one of the quarks inside
the nucleon and the latter by replacing one or several up or down quarks with one or more
strange quarks. The EMFFs of the  Delta and the hyperons provide complementary information about
the intrinsic structure of the nucleon~\cite{Granados:2017cib}.
The electromagnetic properties of the Delta baryon have been studied in detail in
Ref.~\cite{Pascalutsa:2006up}. Recent investigations of the hyperon EM structure are given in
Refs.~\cite{Kubis:2000aa,Haidenbauer:2016won,Granados:2017cib,Alarcon:2017asr,Junker:2019vvy,Haidenbauer:2020wyp,Irshad:2022zga}.
Ref.~\cite{Granados:2017cib} considered once-subtracted dispersion relations for the electromagnetic Sigma-to-Lambda
transition form factors (TFFs) and expressed these in terms of the pion EMFF and the two-pion-Sigma-Lambda scattering
amplitudes. Using an Omn\`{e}s representation, the pion EMFF could be expressed as the Omn\`{e}s function
of the pion $P$-wave phase shift which has been well determined from the Roy-type analyses of the
pion-pion scattering amplitude~\cite{Garcia-Martin:2011iqs}. An improved parameterization of the pion EMFF
is also available, which includes further inelasticities and is applicable at higher energies~\cite{Hanhart:2012wi}.
Moreover, the two-pion-Sigma-Lambda scattering amplitudes could be calculated in a model-independent way
by using  three-flavor chiral perturbation theory (ChPT)~\cite{Kubis:2000aa}. Combining these studies
and taking some reasonable values for couplings and the low-energy constants in three-flavor ChPT,
the electromagnetic Sigma-to-Lambda TFFs were predicted in Ref.~\cite{Granados:2017cib}
where the pion rescattering and the role of the explicit inclusion of the decuplet baryons in three-flavor ChPT
were also investigated.

In the present work, we extend the theoretical framework used in Ref.~\cite{Granados:2017cib}
to explore the effect of the $K\bar{K}$ inelasticity on the electromagnetic Sigma-to-Lambda transition form factors. This is performed by considering the two-channel Muskhelishvili-Omn\`{e}s representation when introducing the pion rescattering effects. In principle, one should include even more inelasticities when
implementing the dispersion theoretical parameterization for the Sigma-to-Lambda TFFs, as done in our previous work
on the nucleon EMFFs~\cite{Lin:2021xrc}. However, it is difficult in the current case due to the
poor data base which is required when constructing reliable inelasticities in the higher energy region, that is above
the $K\bar{K}$ threshold at $\sim 1$~GeV. Note that the four-pion channel has negligible effects in the energy region around $1~\rm GeV$,
see Ref.~\cite{Bijnens:2021hpq}. It is also known that the contribution of the four-pion channel
to the pion and kaon form factors below  $1~\rm GeV$ is a three-loop effect in
ChPT~\cite{Gasser:1990bv} and thus is heavily
suppressed.

We remark that the $4\pi$ channel was shown to play an important role starting from $1.4$~GeV in the $S$-wave case~\cite{Moussallam:1999aq}. This is caused by the presence of the nearby scalar resonances $f_0(1370)$ and $f_0(1500)$ which were both observed to have a sizable coupling to four-pion states~\cite{Adamo:1993bfv,CrystalBarrel:1994doj,CrystalBarrel:1996wfh}. There is no evidence, however, for the presence of corresponding $1^-$ isovector states in the energy region of $1...2$~GeV in the $P$-wave case. Moreover, another experimental finding of these references is that the $4\pi$ system likes to cluster into two resonances in the energy region above 1~GeV~\cite{Moussallam:1999aq}. The lowest candidate is supposed to be the $\rho\rho$ channel for the $P$-wave isovector problem.
From the phenomenological point of view, the inelasticity around 1~GeV should be saturated to a good approximation by the $\pi\pi$ and $K\bar{K}$ coupled-channel treatments in the $P$-wave case.
Moreover, as we will show later, the effect of $K\bar{K}$ inelasticity is small. Thus, the relative ratio
between the effects of the $K\bar{K}$ and four-pion channel could be enhanced. To investigate this
relative ratio, a sophisticated calculation on the four-pion inelasticity is needed which goes
beyond the present work.   

In the present work, the $K\bar{K}$ inelasticity is implemented using SU(3) ChPT. The inclusion of the $K\bar{K}$ channel allows one to construct the Sigma-to-Lambda transition form factors up to $1$~GeV precisely. In addition,
the estimation of the theoretical uncertainties is improved by using the bootstrap approach~\cite{Efron:1993}.

The paper is organized as follows: In Sect.~\ref{sec:theory} we introduce the dispersion theoretical description
of the electromagnetic Sigma-to-Lambda transition form factors and present the coupled-channel
Muskhelishvili-Omn\`{e}s representation for the inclusion of the $K\bar{K}$ inelasticity.
Numerical results are collected in Sect.~\ref{sec:result}. The paper closes with a summary.
Some technicalities are relegated to the appendix.

\section{Formalism}
\label{sec:theory}

Here, we discuss the basic formalism underlying our calculations. We first write down
once-subtracted dispersion relations for the electric and magnetic Sigma-to-Lambda transition form factor
and then discuss in detail their various ingredients, namely the vector form factor of the pion and the
kaon and the amplitudes for $\Sigma^0\bar{\Lambda}\to \pi\pi$ and $\Sigma^0\bar{\Lambda}\to K\bar{K}$,
in order.

\subsection{Dispersion relations for the Sigma-to-Lambda TFFs}
The electromagnetic Sigma-to-Lambda TFFs are defined as in Refs.~\cite{Kubis:2000aa,Granados:2017cib},
\begin{align}
\langle &\Sigma^0(p^\prime) \vert j^\mu \vert \Lambda(p) \rangle \notag\\
&= e \, \bar u(p^\prime) \, \bigg(\left( \gamma^\mu + \frac{m_\Lambda-m_{\Sigma^0}}{t} \, q^\mu \right) \, F_1(t)\nonumber \\ 
& \phantom{mmmmmmmm}  + \frac{i \sigma^{\mu\nu} \, q_\nu}{m_\Lambda + m_{\Sigma^0}} \, F_2(t) \bigg) \, u(p)~,
\label{eq:defFF}
\end{align}
with $t=(p^\prime-p)^2=q^2$ the four-momentum transfer squared. The scalar functions  $F_{1}(t)$ and
$F_{2}(t)$ are called the Dirac and Pauli transition form factors, respectively.
One also writes the electric and
magnetic Sachs transition form factors, given by the following linear combinations,
\begin{align}
	G_E(t)&=F_1(t)+\frac{t}{(m_\Lambda + m_{\Sigma^0})^2}F_2(t),\notag\\
	G_M(t)&=F_1(t)+F_2(t),
\end{align}
with the normalizations $F_1(0)=G_E(0)=0$ and $F_2(0)=G_M(0)=\kappa\approx 1.98$. Here, $\kappa$ is estimated
from the experimental width of the decay $\Sigma^0\to \Lambda\gamma$, see Ref.~\cite{Granados:2017cib} for details.
Unlike the nucleon case where one constructs  dispersion relations for $F_1$ and $F_2$~\cite{Lin:2021umz}, we work
with the electric and magnetic Sachs form factors, i.e. $G_E$ and $G_M$, for the Sigma-to-Lambda TFFs as in
Ref.~\cite{Granados:2017cib} since the Sigma-to-Lambda TFFs are of pure isovector type and the helicity
decomposition used in Ref.~\cite{Granados:2017cib} can easier be applied to the Sachs FFs.

\begin{figure}[t]
	\centerline{\includegraphics*[width=0.35\textwidth,angle=0]{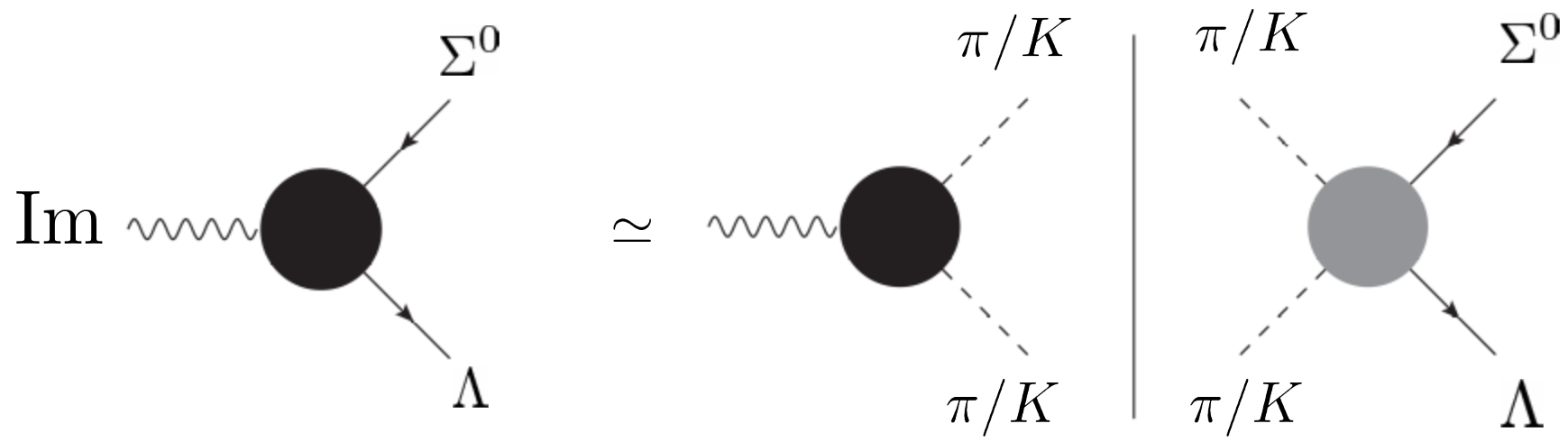}}
	\caption{
		The spectral decomposition of the matrix element of the electromagnetic current $j_\mu$ in Eq.~\eqref{eq:defFFtl}.
	}
	\label{fig:spec}
	\vspace{-3mm}
\end{figure}

In order to apply the spectral decomposition to estimate the imaginary part ${\rm Im}\, G_{E/M}$, we consider
the matrix element of the electromagnetic current Eq.~\eqref{eq:defFF} in the time-like region ($t>0$), which
is obtained via crossing symmetry,
\begin{align}
	\langle &0 \vert j^\mu \vert \Sigma^0(p_3)\bar{\Lambda}(p_4) \rangle \notag\\
	&= e \, \bar{v}(p_4)  \, \bigg(\left( \gamma^\mu + \frac{m_\Lambda-m_{\Sigma^0}}{t} \, (p_3+p_4)^\mu \right)
        \, F_1(t)\nonumber \\ 
	& \phantom{mmmmm}  - \frac{i \sigma^{\mu\nu} \, (p_3+p_4)_\nu}{m_\Lambda + m_{\Sigma^0}} \, F_2(t) \bigg) \, u(p_3)
	\label{eq:defFFtl}
\end{align}
where $p_3$ and $p_4$ are the momenta of the $\Sigma^0$ and $\bar{\Lambda}$ created by the electromagnetic current,
respectively. The four-momentum transfer squared in the time-like region is then $t=(p_3+p_4)^2$. With the
$\pi\pi$ and $K\bar{K}$ inelasticities taken into account as depicted in Fig.~\ref{fig:spec}, the
unitarity relations for the Sigma-to-Lambda TFFs read~\cite{Koepp:1974da,Schneider:2012ez,Granados:2017cib,Junker:2019vvy},
\begin{align}
	\label{eq:unitarity_relation}
	      &\frac1{2i}{\rm disc}_{\rm unit}~G_{E/M}(t)=\frac{1}{12\pi\sqrt{t}} \notag\\
	      &\phantom{mm}\times \bigg( q_\pi^3(t)\,F_\pi^V(t)^*\, T_{E/M}^{\pi\pi}(t)\,
              \theta\left(t-4M_\pi^2\right)~\notag\\
	&\phantom{mm}+2{q_K^3(t)} \,F_K^V(t)^*\, T_{E/M}^{K\bar{K}}(t)\, \theta\left(t-4M_K^2\right)\bigg)~,
\end{align}
where
\begin{equation}\label{eq:qcm}
q_{\pi/K}(t)=\sqrt{\frac{\lambda(M_{\pi/K}^2,M_{\pi/K}^2,t )}{4t}}
\end{equation}
is the center-of-mass momentum of the $\pi\pi/K\bar{K}$ two-body continuum with
$\lambda(x,y,z)=x^2+y^2+z^2-2(xy+yz+zx)$ the K\"{a}ll\'{e}n function. $F_{\pi/K}^V(t)$ is the vector-isovector form
factor ($J=I=1$) of the pion/kaon and $T_{E}^{\pi\pi}(t)$ and $T_{M}^{\pi\pi}(t)$ are two independent reduced $P$-wave
$\Sigma^0\bar\Lambda\to \pi^+\pi^-$ amplitudes in the helicity basis. Similarly, $T_{E}^{K\bar{K}}(t)$ and
$T_{M}^{K\bar{K}}(t)$ denote the corresponding reduced amplitudes for $\Sigma^0\bar\Lambda\to K^+K^-(K^0\bar{K}^0)$. Then
the once-subtracted dispersion relations for the Sigma-to-Lambda TFFs are written as,
\begin{align}
	\label{eq:tffDR}
	G_{E/M}(t)&=G_{E/M}(0)+\frac{t}{2\pi i}\int_{4M_\pi^2}^\infty d t^\prime \frac{{\rm disc}_{\rm unit}\, G_{E/M}(t^\prime)}{t^\prime(t^\prime-t-i\epsilon)}\notag\\
	&\phantom{mmmm}+G^{\rm anom}_{E/M}(t)~,
\end{align}
where the last term $G^{\rm anom}_{E/M}(t)$ denotes the contribution of the anomalous cut which is non-zero when there exists 
an anomalous threshold in the involved processes~\cite{Karplus:1958zz,Lucha:2006vc,Hoferichter:2013ama,Molnar:2019uos,Junker:2019vvy}. This does happen when the $K\bar{K}$ channel is taken into account, see Appendix~\ref{app:app1} for detailed discussions and the explicit expressions of the anomalous part. Next, we need to consider the various factors contributing to Eq.~\eqref{eq:tffDR}.

\subsection{The $\Sigma^0\bar\Lambda\text{-}\pi\pi$ and $\Sigma^0\bar\Lambda\text{-}K\bar{K}$ amplitudes in the two-channel Muskhelishvili-Omn\`{e}s representation}
We start with the four-point amplitudes
$\Sigma^0\bar\Lambda\to\pi^+\pi^-$ and $\Sigma^0\bar\Lambda\to K^+K^-$. Note that the matrix element
Eq.~\eqref{eq:defFFtl}, and also the four-point function $\Sigma^0\bar\Lambda \to\pi\pi$, can be written
in the general form $\bar{v}_{\Lambda}(-p_z,\lambda)\Gamma u_{\Sigma^0}(p_z,\sigma)$ when one works in the
center-of-mass frame and chooses the z-axis along the direction of motion of the $\Sigma^0$.
Here, $\sigma$ and $\lambda$ are the helicities of the $\Sigma^0$ and $\bar{\Lambda}$ baryons, respectively.
Due to parity invariance, there are only two non-vanishing terms, $\sigma=\lambda=+1/2$ and $\sigma=-\lambda=+1/2$.
Concerning the matrix element Eq.~\eqref{eq:defFFtl}, all components
except for $\mu=3$ vanish in the case of $\sigma=\lambda=+1/2$, that is,
\begin{align}
	\label{eq:nonzeroGE}
	\langle &0 \vert j^3 \vert \Sigma^0(p_z,\frac12)\bar{\Lambda}(-p_z,\frac12) \rangle\notag\\
	&\phantom{mmmm}=\bar{v}_{\Lambda}(-p_z,+1/2)\gamma^3 u_\Sigma(p_z,+1/2)G_E(t).
\end{align}
For $\sigma=-\lambda=+1/2$, only components related to $\mu=1,2$ survive:
\begin{align}
	\label{eq:nonzeroGM}
	\langle &0 \vert j^1 \vert \Sigma^0(p_z,\frac12)\bar{\Lambda}(-p_z,-\frac12) \rangle\notag\\
	&=\bar{v}_{\Lambda}(-p_z,-1/2)\gamma^1 u_\Sigma(p_z,+1/2)G_M(t),
\end{align}
and the matrix element for $\mu=2$ differs from $\mu=1$ only by a factor $i$. Eqs.~\eqref{eq:nonzeroGE} and \eqref{eq:nonzeroGM}
show that $T_{E}$ in the imaginary part of $G_E$ is only related to the amplitude component ${\cal M}^{\Sigma\bar\Lambda\to\pi\pi/{K\bar{K}}}(+1/2,+1/2)$
while $T_{M}$ comes from ${\cal M}^{ \Sigma\bar\Lambda\to\pi\pi/{K\bar{K}}}(+1/2,-1/2)$. Then we define the reduced amplitudes $T_{E/M}$
as~\cite{Jacob:1959at,Granados:2017cib}
\begin{align}
	&{\cal M}^{\Sigma\bar\Lambda\to\pi\pi/{K\bar{K}}}(t,\theta,+1/2,+1/2)=\notag\\
	&\phantom{mm}\bar{v}_{\Lambda}(-p_z,+1/2)\gamma^3u_\Sigma(p_z,+1/2)q_{\pi/K}T_E^{\pi/K}(t)d_{0,0}^1(\theta)\notag\\
	&\phantom{mm}+\text{other partial waves with}~J\neq 1~,\\
	&{\cal M}^{\Sigma\bar\Lambda\to\pi\pi/{K\bar{K}}}(t,\theta,+1/2,-1/2)=\notag\\
	&\phantom{m}-\sqrt{2}\bar{v}_{\Lambda}(-p_z,-1/2)\gamma^1u_\Sigma(p_z,+1/2)q_{\pi/K}T_M^{\pi/K}(t)d_{1,0}^1(\theta)\notag\\
	&\phantom{mm}+\text{other partial waves with}~J\neq 1~,
\end{align} 
where $d^{1}_{1/2\pm 1/2,0}(\theta)$ is the Wigner $d$-matrix. Finally, we obtain
\begin{align}
	\label{eq:reduced_ampE}
	&T_E^{\pi/K}(t)=\notag\\
	&\frac32 \, \int\limits_0^\pi d\theta \, \sin\theta \, 
	\frac{{\cal M}^{\Sigma\bar\Lambda\to\pi\pi/{K\bar{K}}}(t,\theta,+1/2,+1/2)}{\bar v_\Lambda(-p_z,+1/2) \, \gamma^3 \, u_\Sigma(p_z,+1/2) \; q_{\pi/K}} 
	\, \cos\theta~, \\
	\label{eq:reduced_ampM}
	&T_M^{\pi/K}(t)=\notag\\
	&\frac34 \, \int\limits_0^\pi d\theta \, \sin\theta \, 
	\frac{{\cal M}^{\Sigma\bar\Lambda\to\pi\pi/{K\bar{K}}}(t,\theta,+1/2,-1/2)}{\bar v_\Lambda(-p_z,-1/2) \, \gamma^1 \, u_\Sigma(p_z,+1/2) \; q_{\pi/K}} 
	\, \sin\theta~.
\end{align}

\begin{figure}[t]
	\centerline{\includegraphics*[width=0.35\textwidth,angle=0]{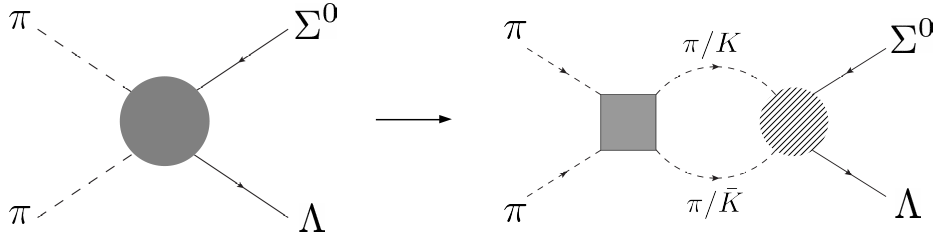}}
	\caption{
	  The four-point function $\pi\pi\to\Sigma^0\bar\Lambda$  including two-pion rescattering.
          The hatched circle is the part containing only left-hand cuts and a polynomial. 
	}
	\label{fig:rescatter}
	\vspace{-3mm}
\end{figure}

As done in Ref,~\cite{Granados:2017cib}, the pion rescattering effect can be introduced into $T_{E/M}$ via the Muskhelishvili-Omn\`{e}s equation that is shown in
Fig.~\ref{fig:rescatter}. With the inclusion of the $K\bar{K}$ channel, the two-channel 
Muskhelishvili-Omn\`{e}s representation reads~\cite{Babelon:1976ww,Morgan:1991zx,Garcia-Martin:2010kyn,Hoferichter:2012wf}
\begin{align}
	\label{eq:coupled-channel}
	&\vec{T}_{E/M}(t)=\vec{K}_{E/M}(t)+\vec{T}_{E/M}^{\rm anom}(t)\notag\\
	&+\vec\Omega(t)\left(\vec{P}_{0,E/M}(t)-\frac{t}{\pi}\int_{4M_{\pi}^2}^\infty {d t^\prime}\frac{[ {\rm Im} \vec\Omega^{-1}(t^\prime) ]\vec{K}_{E/M}(t^\prime)}{t^\prime(t^\prime-t-i\epsilon)}\right)~,
\end{align}
with $\vec{T}=\begin{pmatrix} T_\pi, & \sqrt{2}T_K\end{pmatrix}^T$, $\vec{P}_0=\begin{pmatrix} P_{0,\pi}, & \sqrt{2}P_{0,K}\end{pmatrix}^T$ and $\vec{K}=\begin{pmatrix} K_\pi, & \sqrt{2}K_K\end{pmatrix}^T$.\footnote{ $\vec{K}=\begin{pmatrix} K_\pi\theta(t^\prime-4M_\pi^2), & \sqrt{2}K_K\theta(t^\prime-4M_K^2)\end{pmatrix}^T$ is implicit in the integrand $[ {\rm Im} \vec\Omega^{-1}(t^\prime) ]\vec{K}(t^\prime)$.} $\vec\Omega(t)$ is the two-dimension Omn\`{e}s matrix for the $P$-wave isovector $\pi\pi$-$K\bar{K}$ coupled-channel system. $\vec{T}_{E/M}^{\rm anom}(t)$ again indicates the anomalous contribution and its explicit formula is given in Appendix.~\ref{app:app1}. The $K_{\pi/K}$ denotes the part of the
$\Sigma\bar\Lambda\to\pi\pi/{K\bar{K}}$ amplitude that only contains the left-hand cut (LHC) and $P_0$ is the remainder which is purely polynomial.
Therefore, $K_{\pi/K}$ and $P_{0, \pi/K}$ are given by
\begin{align}
	\label{eq:kE}
	&K_E(t)=\notag\\
	&\frac32 \, \int\limits_0^\pi d\theta \, \sin\theta \, 
	\frac{{\cal M}^{\rm pole}(t,\theta,+1/2,+1/2)}{\bar v_\Lambda(-p_z,+1/2) \, \gamma^3 \, u_\Sigma(p_z,+1/2) \; q_{\pi/K}} 
	\, \cos\theta~, \\
	\label{eq:pE}
	&P_0^E(t)=\notag\\
	&\frac32 \, \int\limits_0^\pi d\theta \, \sin\theta \, 
	\frac{{\cal M}^{\rm contact}(t,\theta,+1/2,+1/2)}{\bar v_\Lambda(-p_z,+1/2) \, \gamma^3 \, u_\Sigma(p_z,+1/2) \; q_{\pi/K}} 
	\, \cos\theta~,
\end{align}
with ${\cal M}^{\Sigma\bar\Lambda\to\pi\pi/{K\bar{K}}}(t,\theta)={\cal M}^{\rm pole}+{\cal M}^{\rm contact}$. The magnetic parts are derived equivalently from Eq.~\eqref{eq:reduced_ampM},
\begin{align}
	\label{eq:kM}
	&K_M(t)=\notag\\
	&\frac34 \, \int\limits_0^\pi d\theta \, \sin\theta \, 
	\frac{{\cal M}^{\rm pole}(t,\theta,+1/2,-1/2)}{\bar v_\Lambda(-p_z,-1/2) \, \gamma^1 \, u_\Sigma(p_z,+1/2) \; q_{\pi/K}} 
	\, \sin\theta~, \\
	\label{eq:pM}
	&P_0^M(t)=\notag\\
	&\frac34 \, \int\limits_0^\pi d\theta \, \sin\theta \, 
	\frac{{\cal M}^{\rm contact}(t,\theta,+1/2,-1/2)}{\bar v_\Lambda(-p_z,-1/2) \, \gamma^1 \, u_\Sigma(p_z,+1/2) \; q_{\pi/K}} 
	\, \sin\theta~.
\end{align}
All the reduced amplitudes $K_{\pi/K}$ and $P_{0, \pi/K}$ are calculated up to next-to-leading order (NLO) within the framework of the three-flavor baryon ChPT that includes the decuplet baryon as explicit degrees of freedom. Their explicit expressions are derived in detail in Appendix~\ref{app:chpt}.

\subsection{The $P$-wave Omn\`{e}s matrix and $\pi$, $K$ vector form factors}
In this subsection, we derive the $P$-wave isovector Omn\`{e}s matrix and solve for the pion and kaon EMFFs in the coupled-channel formalism. $\vec\Omega$ satisfies the unitarity relation~\cite{Hoferichter:2012wf}
\begin{equation}
	\label{eq:omnesim}
	\frac1{2i}{\rm disc}\ \Omega_{ij}=(t_1^1)^*_{im}\Sigma_m^{}\Omega_{mj}^{}~,
\end{equation}
where 
\begin{equation}
	\vec\Sigma(t)={\rm diag}\left(\sigma_\pi q_\pi^{2}\theta(t-4M_\pi^2),\sigma_K q_K^{2}\theta(t-4M_K^2)\right)~
\end{equation}
with 
\begin{equation}\label{eq:sigma}
	\sigma_{\pi/K}(t)=\sqrt{1-\frac{4 M_{\pi/K}^2}{t}}
\end{equation}
is the diagonal phase space matrix. The $J=I=1$ $\pi\pi$-$K\bar{K}$ coupled-channel $T$-matrix $\vec t^1_1$ is  parameterized as
\begin{equation}
	\label{eq:coupledT}
	\vec t^1_1=\begin{pmatrix} \displaystyle\frac{\eta e^{2i \delta^1_1}-1}{2i \sigma_\pi q_\pi^{2}}& g e^{i\psi} \\ g e^{i\psi} & \displaystyle\frac{\eta e^{2i (\psi-\delta^1_1)}-1}{2i \sigma_K q_K^{2}}\end{pmatrix}~.
\end{equation}
where $g$ and $\psi$ are the modulus and phase of the $P$-wave isovector $\pi\pi\to K\bar{K}$ scattering amplitude, respectively. The inelasticity $\eta$ is defined by
\begin{equation}
	\eta(t)=\sqrt{1-4\sigma_\pi\sigma_K(q_\pi q_K)^2 g^2\theta(t-4M_K^2)}.
\end{equation}
Then we can write the dispersion relation for the Omn\`{e}s matrix $\vec\Omega$ as
\begin{equation}
	\label{eq:MO}
	\Omega_{ij}(t)=\frac1{2\pi i}\int_{4M_\pi^2}^\infty {\rm d} z\frac{{\rm disc}\ \Omega_{ij}(z)}{z-t-i\epsilon}~.
\end{equation}
The analytic solution of the integral equation Eq.~\eqref{eq:MO} was given in Ref.~\cite{Omnes:1958hv}
for the single-channel problem.
However, there are no known analytic solutions for two or more
channel cases where one has to construct the solutions numerically, either by an iterative
procedure~\cite{Donoghue:1990xh} or a discretization method~\cite{Moussallam:1999aq}. Here,
we adopt the iterative approach to solve the $P$-wave $\pi\pi$-$K\bar{K}$ coupled-channel Omn\`{e}s matrix.
Substituting Eq.~\eqref{eq:omnesim} into Eq.~\eqref{eq:MO}, one obtains a two-dimensional system of
integral equations
\begin{equation}\label{eq:omsolve}
 \begin{cases}
   {\rm Re}\ \chi_1(t)=\displaystyle\frac1{\pi}\mathcal{P}\displaystyle\int_{4M_\pi^2}^\infty {\rm d} z\displaystyle
   \frac{{\rm Im}\chi_1(z)}{z-t}~,\\
        {\rm Re}\ \chi_2(t)=\displaystyle\frac1{\pi}\mathcal{P}\displaystyle\int_{4M_\pi^2}^\infty {\rm d} z\displaystyle
        \frac{{\rm Im}\chi_2(z)}{z-t}~,
	\end{cases} 
\end{equation}
where
\begin{align}
  {\rm Im}\chi_1(z)&= {\rm Re}\left[(t_1^1)^*_{11}\Sigma_1^{} \chi_1^{}\right]+{\rm Re}\left[(t_1^1)^*_{12}\Sigma_2^{}
    \chi_2^{}\right]~, \notag\\
  {\rm Im}\chi_2(z)&= {\rm Re}\left[(t_1^1)^*_{21}\Sigma_1^{} \chi_1^{}\right]+{\rm Re}\left[(t_1^1)^*_{22}\Sigma_2^{}
    \chi_2^{}\right]~,
\end{align}
and $\mathcal{P}$ denotes the principal value.
Searching for solutions of $\vec{\Omega}(t)$ is equivalent to searching for two independent
solutions of the integral equation set for the two-dimensional array $(\chi_1,\chi_2)^T$. Using the
iterative procedure, one can obtain a series of solutions $(\chi_1^\lambda,\chi_2^\lambda)^T$ starting
with various initial inputs $\chi_1(t)=1$, $\chi_2(t)=\lambda$, where $\lambda$ is a real
parameter. Note that the iterative process is linear and the results of the iteration is
therefore a linear function of $\lambda$~\cite{Donoghue:1990xh}. Then the solution family
$\{(\chi_1^\lambda,\chi_2^\lambda)^T\}$ contains only two linearly independent members.
Here, we take the same convention as Ref.~\cite{Hoferichter:2012wf} to construct two independent
solutions, $(\Omega_{11}, \Omega_{21})^T$ and $(\Omega_{12}, \Omega_{22})^T$, that satisfy the normalizations
\begin{equation}
	 \Omega_{11}(0)=\Omega_{22}(0)=1~~\text{and}~~\Omega_{12}(0)=\Omega_{21}(0)=0~,\notag
\end{equation}
from two arbitrary solutions $(\chi_1^{\lambda_1},\chi_2^{\lambda_1})^T$ and $(\chi_1^{\lambda_2},\chi_2^{\lambda_2})^T$.

With the two-channel Muskhelishvili-Omn\`{e}s representation, the binary function composed of the vector FFs of the pion and the kaon fulfills the same unitarity relation Eq.~\eqref{eq:omnesim}. Then one can solve the pion and kaon vector form factors 
\begin{equation}\label{eq:isoff}
	\begin{pmatrix} F_\pi^V(t) \\ \sqrt{2}F_K^V(t)\end{pmatrix}=\begin{pmatrix} \Omega_{11}(t) & \Omega_{12}(t) \\ \Omega_{21}(t) & \Omega_{22}(t)\end{pmatrix}\begin{pmatrix} F_\pi^V(0) \\ \sqrt{2}F_K^V(0)\end{pmatrix},
\end{equation}
which are normalized as $F_\pi^V(0)=1$ and $F_K^V(0)=1/2$.

To solve the $J=I=1$ $\pi\pi$-$K\bar{K}$ Omn\`{e}s matrix, the required input is
the $P$-wave isovector $\pi\pi$-$K\bar{K}$ scattering matrix $\vec{t}_1^1$, i.e. Eq.~\eqref{eq:coupledT}, that is constructed from the $\pi\pi$ $P$-wave isovector phase shift $\delta^1_1$, the modulus $g$ and phase $\psi$ of the $P$-wave isovector $\pi\pi\to K\bar{K}$ amplitude. The phase shift $\delta_1^1$ up to $1.4~{\rm GeV}$ was
extracted precisely from the Roy-type analyses of the pion-pion scattering amplitude in Ref.~\cite{Garcia-Martin:2011iqs}. 
We take the same prescription as in Ref.~\cite{Hanhart:2012wi} to
extrapolate it smoothly to reach $\pi$ at infinity. Then $\delta_1^1(t)$ is given by
\begin{equation}
	\label{eq:deltaP}
	\delta_1^1(t)=
	\begin{cases}
		0, & 0\le \sqrt{t}\le 2M_\pi,\\
		\delta_{f_1}(t), & 2M_\pi < \sqrt{t}\le 2M_K,\\
		\delta_{f_2}(t), & 2M_K < \sqrt{t}\le \sqrt{t_0},\\
		\delta_{f_3}(t), & \sqrt{t_0}< \sqrt{t},
	\end{cases}
\end{equation}
where
\begin{align}
	\delta_{f_1}(t)&=\cot^{-1}\Bigg(\frac{\sqrt{t}}{2q_\pi^3}(M_\rho^2-t)\bigg(\frac{2M_\pi^3}{M_\rho^2\sqrt{t}}
	+1.043~\notag\\
	&\phantom{mmmmmmm}+0.19\frac{\sqrt{t}-\sqrt{t_1-t}}{\sqrt{t}+\sqrt{t_1-t}}
	\bigg)\Bigg)~,\notag\\
	\delta_{f_2}(t)&=\delta_{f_1}(4M_K^2)+1.39\left(\frac{\sqrt{t}}{2M_K}-1\right)-1.7\left(\frac{\sqrt{t}}{2M_K}-1\right)^2~,\notag\\
	\delta_{f_3}(t)&=\pi+(\delta_{f_2}(t_0)-\pi) \left(\frac{t_2+t_0}{t_2+t}\right)~.
\end{align}
Here, $t_0=(1.4~{\rm GeV})^2$, $t_1=(1.05~{\rm GeV})^2$ and $t_2=(10~{\rm GeV})^2$. 
\begin{figure}[t]
	\centerline{\includegraphics*[width=0.45\textwidth,angle=0]{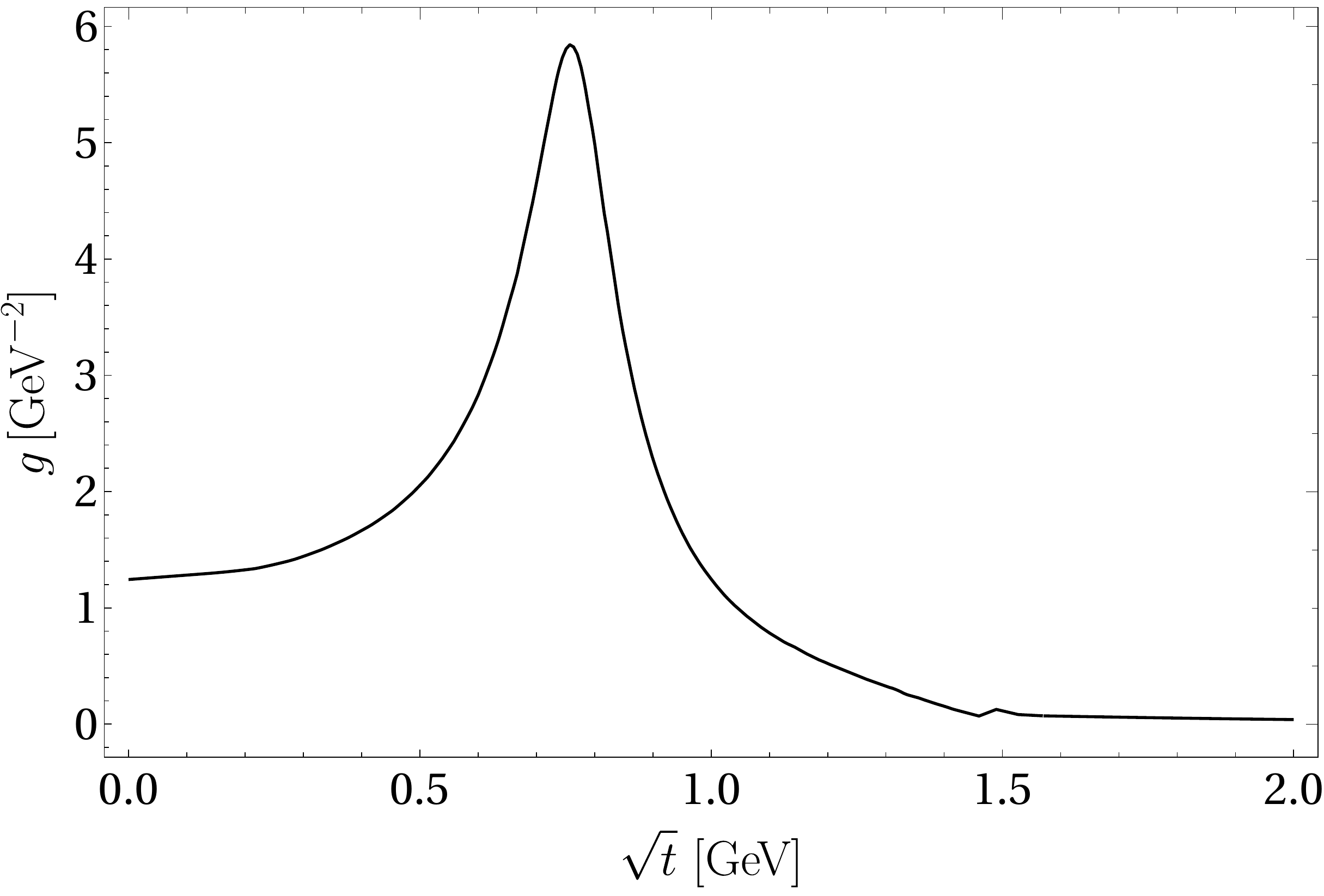}}
	\caption{
		The modulus $g$ of the $P$-wave $\pi\pi\to K\bar{K}$ amplitude given by Ref.~\cite{Buettiker:2003pp}.
	}
	\label{fig:modulus}
	\vspace{-3mm}
\end{figure} 
The $P$-wave $\pi\pi\to K\bar{K}$ amplitude up to $\sqrt{t_3}=1.57~{\rm GeV}$
is taken from Ref.~\cite{Buettiker:2003pp} where the modulus $g$ in the region of $4M_\pi^2$...$4M_K^2$ was solved from the Roy-Steiner equation with the experimental data of $P$-wave $\pi\pi\to K\bar{K}$ scattering~\cite{Cohen:1980cq,Etkin:1981sg} above the $K\bar{K}$ threshold as input, while the phase $\psi$ was fitted to experimental data~\cite{Cohen:1980cq,Etkin:1981sg}.
Note that the two-channel Muskhelishvili-Omn\`{e}s representation in terms of $\pi\pi$ and
$K\bar{K}$ intermediate states should only work well in the lower energy region~\cite{Hoferichter:2012wf}. Further, the asymptotic values of phase shifts in the coupled-channel systems have to satisfy
\begin{equation}
	\lim_{t\to\infty}\sum \delta^I_l(t)\geq n\pi,
\end{equation}
to ensure that the system of integral equations, Eq.~\eqref{eq:omsolve}, has a unique solution ~\cite{Moussallam:1999aq,Yao:2018tqn}. $n$ is the number of channels that are considered in the formalism. It requires $\psi=\delta_{1,\pi\pi}^1+\delta_{1,K\bar{K}}^1\geq 2\pi$ in Eq.~\eqref{eq:coupledT}. $g$ and $\psi$ are extrapolated smoothly to $0$ 
and $2\pi$ by means of~\cite{Moussallam:1999aq}
\begin{align}
	\psi(t)&=2\pi+(\psi(t_4)-2\pi)\hat{f}\left(\frac t{t_4}\right),\notag\\
	g(t)&=g(t_3)\hat{f}\left(\frac t{t_3}\right),\quad \text{with}\, \hat{f}(x)=\frac{2}{1+x^{3/2}}.
\end{align}
where the extrapolation point $t_4$ of $\psi$ should be far away from $1.5$~GeV since there is a structure located around $1.5$~GeV in the phase of the $P$-wave $\pi\pi\to K\bar{K}$ amplitude. 
Here we take the value $\sqrt{t_4}=5$~GeV for $\psi$. 
Such a structure should also leave trails in the modulus $g$.
However, only $g$ up to $\sqrt{2}~{\rm GeV}$ is estimated in Ref.~\cite{Buettiker:2003pp} and a small bump around $1.5$~GeV in $g$ is only reflected roughly by several data points above $1.4$~GeV measured by Ref.~\cite{Cohen:1980cq}, see Fig. 9 in Ref.~\cite{Buettiker:2003pp}. The modulus used in our work is presented in Fig.~\ref{fig:modulus}, while the $\delta^1_1$ and
$\psi$ are presented when we show the solved pion and kaon vector form factors. 

%
\begin{figure*}[t]
	\centering
	\includegraphics*[width=0.7\textwidth,angle=0]{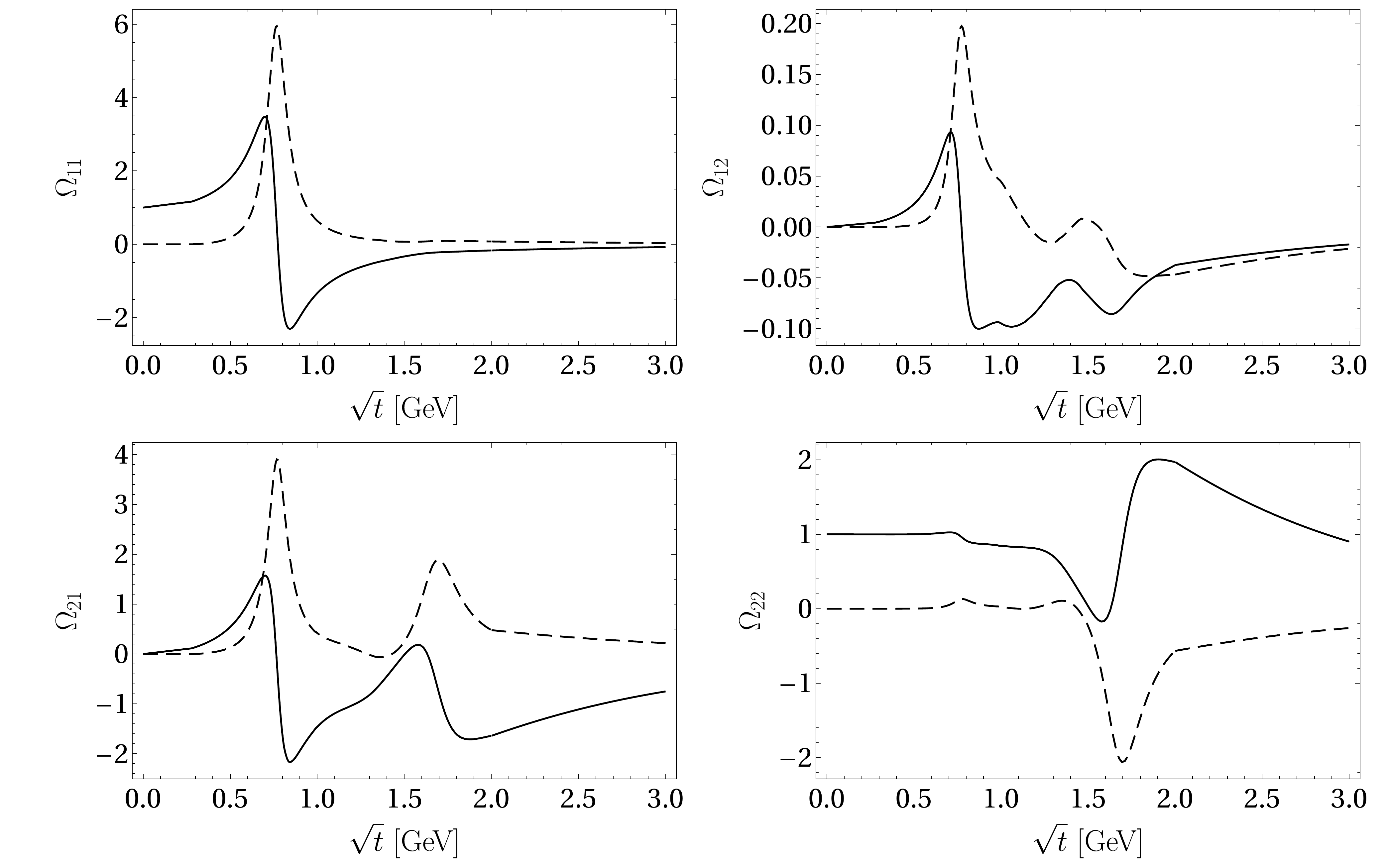}
	\caption{Real (solid line) and imaginary (dashed line) parts of the Omn\`{e}s matrix elements $\Omega$. }
	\label{fig:Omnes}
	\vspace{-3mm}
\end{figure*}
%
%
\begin{widetext}
\begin{figure}[htbp]
	\centering
	\includegraphics*[width=0.7\textwidth,angle=0]{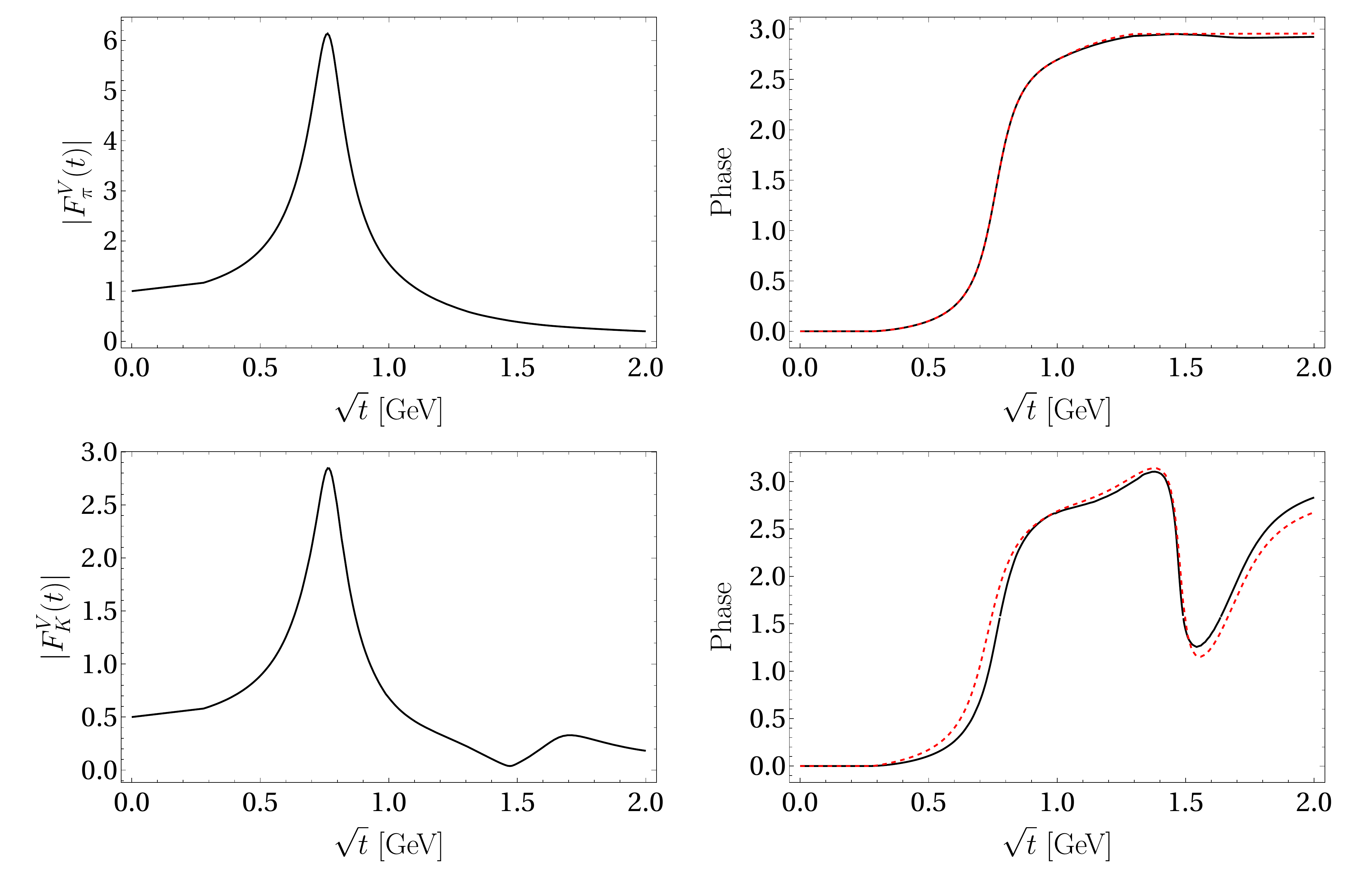}
	\caption{Modulus (left) and phase (right) of the vector pion (upper) and kaon (lower) form factors given by Eq.~\eqref{eq:isoff}. The input $\pi\pi$ phase shift $\delta^1_1$ (upper) and phase $\psi$ (lower) of the $P$-wave isovector $\pi\pi\to K\bar{K}$ amplitude are also presented as the red-dashed lines for comparison. Note that the asymptotic values of $\delta_1^1$ and $\psi$ are $\pi$ and $2\pi$ respectively. The latter is invisible in the plot since its extrapolation point is set as $5~{\rm GeV}$. }
	\label{fig:isoFF}
\end{figure}
\end{widetext}

The obtained $\Omega$ matrix elements are presented in Fig.~\ref{fig:Omnes}.
The pion and kaon vector form factors calculated from Eq.~\eqref{eq:isoff} are then given in Fig.~\ref{fig:isoFF}. Clearly, one can see from Fig.~\ref{fig:isoFF} that the phase of $F_\pi^V$ and $F_K^V$ are consistent with the input $\pi\pi$ phase shift $\delta_1^1$ and the phase $\psi$ of the $P$-wave $\pi\pi\to K\bar{K}$ scattering amplitude respectively, which is similar with the finding for the $S$-wave case by Ref.~\cite{Hoferichter:2012wf}.

\section{Results}\label{sec:result}
Using the reduced amplitudes $P_{0,\pi/K}^{E/M}$ and $K_{\pi/K}^{E/M}$ given in Appendix~\ref{app:chpt} and the anomalous expressions presented in Appendix~\ref{app:app1}, we
can now calculate  the amplitudes $T_{E/M}^{\pi/K}$ in Eq.~\eqref{eq:unitarity_relation}
including $\pi\pi/K\bar{K}$ rescattering effects from  Eq.~\eqref{eq:coupled-channel}. Finally, we calculate the Sigma-to-Lambda transition form factors $G_{E/M}$ from the dispersion relations Eq.~\eqref{eq:tffDR}.
Two issues remain to be clarified. First, we have to fix all the couplings in the expressions of
$P_{0,\pi/K}^{E/M}$ and $K_{\pi/K}^{E/M}$. These are $D$, $F$, $F_\Phi$ for the LO octet-to-octet interactions,
$h_A$ for the LO decuplet-to-octet interaction, and $b_{10}$ for the NLO octet-to-octet interaction.
In ChPT, $D$ and $F$ are well constrained around $0.8$ and $0.5$, respectively. Here we use
$D=0.80$, $F=0.46$~\cite{Kubis:2000aa}. In SU(3) ChPT, $F_\Phi$ can take three different values at LO, namely
$F_\pi=92.4$~MeV, $F_K=113.0$~MeV and $F_\eta=120.1$~MeV~\cite{Mai:2009ce}. Often, one  chooses the average
of these, that is, $F_\Phi=(F_\pi+F_K+F_\eta)/3$. Here, we take $F_\Phi=100\pm 10$~MeV to cover mainly
the $\pi$ and $K$ contributions. $h_A$ can be determined from the experimental widths of either
$\Sigma^*\to\Lambda\pi$ or $\Sigma^*\to\Sigma\pi$. We take the value $h_A=2.3\pm0.3$~\cite{Granados:2017cib}, here an additional 10\% error is added to account for the SU(3) flavor symmetry breaking effect when applied to the vertices involving a $\Xi^*$.
The low-energy constant $b_{10}$
was estimated in Ref.~\cite{Meissner:1997hn} based on the resonance saturation hypothesis as
$b_{10}=0.95~{\rm GeV}^{-1}$. A larger value $b_{10}=1.24~{\rm GeV}^{-1}$ is used in Ref.~\cite{Kubis:2000aa}. A
very recent determination based on the ChPT fits to  lattice data of the axial-vector currents of the
octet baryons gives $b_{10}=0.76~{\rm GeV}^{-1}$~\cite{Sauerwein:2021jxb}. Taking all these
determinations into account, $b_{10}=(1.0\pm0.3)~{\rm GeV}^{-1}$ is used here. Second, we introduce
an energy cutoff $\Lambda$ in the integration along the unitarity cut in Eq.~\eqref{eq:tffDR} and Eq.~\eqref{eq:coupled-channel}.
We consider two values for the cutoff, $\Lambda=1.5$ and $2.0~{\rm GeV}$, to check the sensitivity of our results to it.

\begin{figure}[h]
	\centerline{\includegraphics*[width=0.45\textwidth,angle=0]{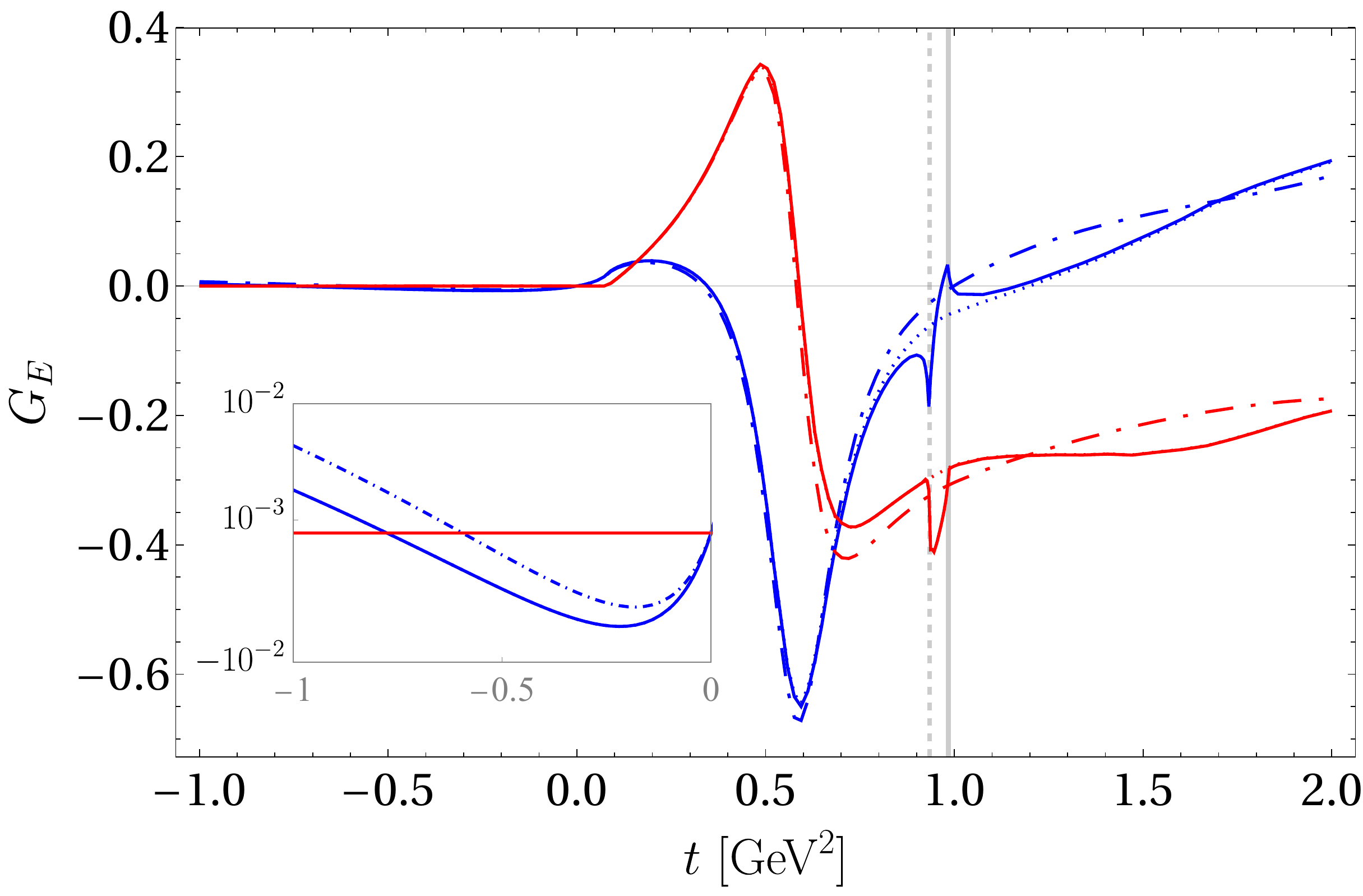}}
	\caption{
		The imaginary (red) and real (blue) part of the electric transition form factor $G_E$. The dash-dotted, dotted, solid lines denote the results within the single $\pi\pi$ channel, $\pi\pi$-$K\bar{K}$ coupled channel without and with the anomalous contribution scenarios, respectively, when $F_\Phi=100$~MeV, $b_{10}=1.06~{\rm GeV^{-1}}$, $h_A=2.22$ and $\Lambda=1.5$~GeV. The vertical dashed and solid lines represent respectively the anomalous threshold (Eq.~\eqref{eq:anompos}) and the $K\bar{K}$ threshold.
	}
	\label{fig:ge0}
	\vspace{-3mm}
\end{figure}

Now we are in the position to present our numerical results for the electromagnetic Sigma-to-Lambda
transition form factors. First, we present the electric transition
form factor $G_E$ obtained with the radius-adjusted parameters given in Ref.~\cite{Granados:2017cib}, i.e. $F_\Phi=100$~MeV, $b_{10}=1.06~{\rm GeV^{-1}}$ and $h_A=2.22$ where the radius is adjusted to the fourth-order ChPT result from Ref.~\cite{Kubis:2000aa}, in Fig.~\ref{fig:ge0}. Note that $\Lambda=1.5$~GeV is used in these calculations. The result from the single $\pi\pi$ channel consideration is also plotted for an intuitive comparison. Taking the same parameter values, we find good agreement with Ref.~\cite{Granados:2017cib}. After the inclusion of the $K\bar{K}$ inelasticity, a logarithmic singularity located at the anomalous threshold $t_-=0.935~{\rm GeV}$ in the unphysical area of the time-like region is introduced into the TFF $G_E$. Moreover, additional nonzero imaginary parts along the anomalous cut are produced for the TFFs by Eq.~\eqref{eq:anomG} and Eq.~\eqref{eq:anomTf}. 
This is similar to the triangle singularity mechanism that leads to a quasi-state phenomenon in the physical observables~\cite{Guo:2019twa}, except the anomalous threshold here can not be accessed directly by the experiments. 
The imaginary parts of $G_E$ in the space-like region, however, are still zero since the nonzero contributions from Eq.~\eqref{eq:anomG} are exactly canceled by those from the unitarity integral of Eq.~\eqref{eq:anomTf}.
\begin{figure}[htbp]
	\centerline{\includegraphics*[width=0.45\textwidth,angle=0]{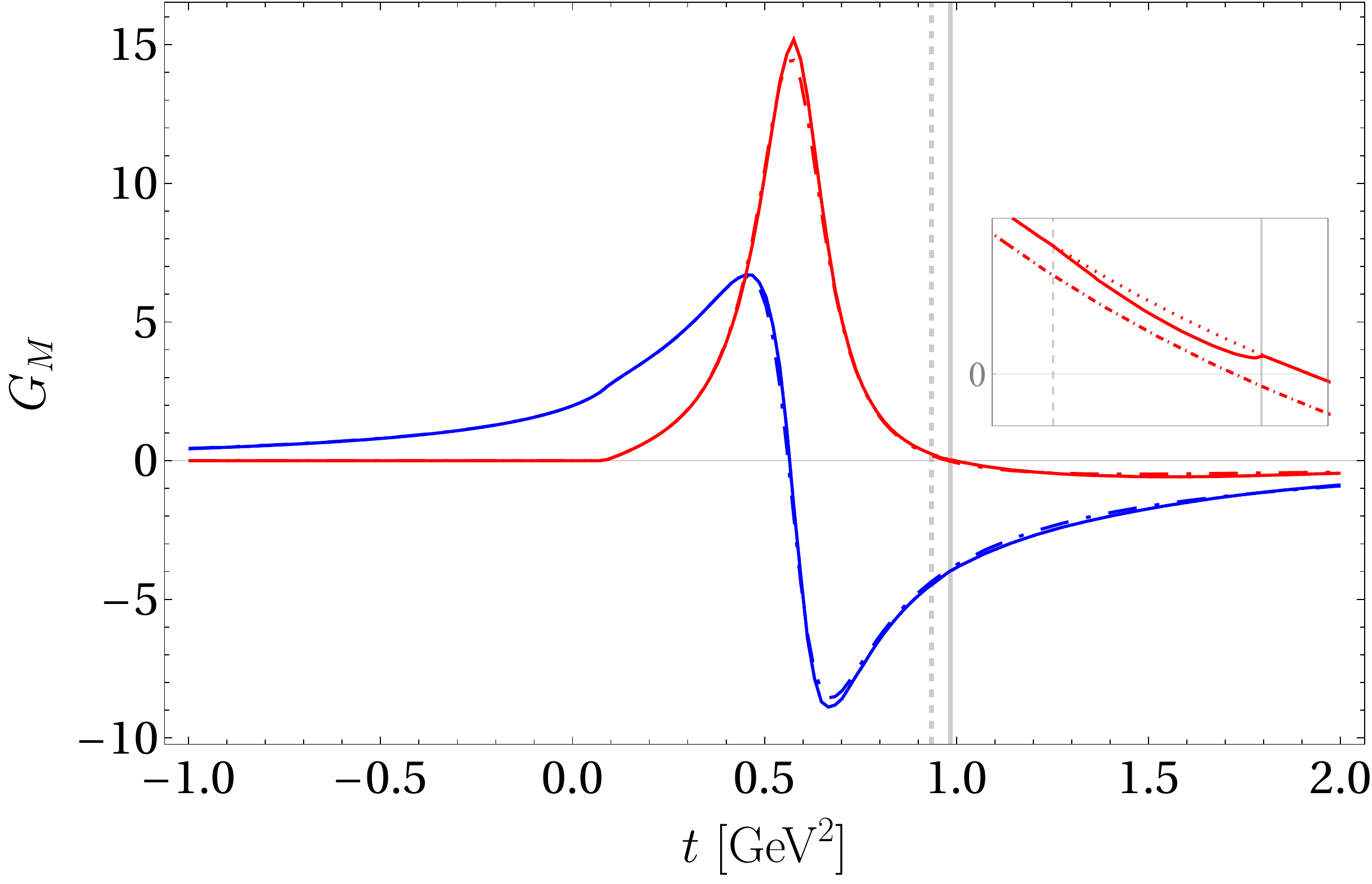}}
	\caption{
		The imaginary (red) and real (blue) part of the magnetic transition form factor $G_M$. For notations, see Fig.~\ref{fig:ge0}.
	}
	\label{fig:gm0}
\end{figure} 
A similar plot for the magnetic TFF $G_M$ is shown in Fig.~\ref{fig:gm0} where there is a cusp-like structure rather than a logarithmic singularity in $G_E$ located at the anomalous threshold since the coefficient $f$ in Eq.~\eqref{eq:anomK} which is proportional to $(Y^2-\kappa^2)$ does vanish at the anomalous threshold for $G_M$. Note that such cusp-like structure is almost invisible due to the large scale variation of the magnitude of $G_M$. With that set of parameters, a $52\%$ decrease is produced by the $K\bar{K}$ channel for $G_E$ at $t=-1~{\rm GeV^2},$\footnote{Note that $G_E$ is overall very small, as is expected due to the vanishing overall charge of the $\Lambda$ and $\Sigma^0$.} while only a $3\%$ decrease happens for $G_M$. One should be aware, however, of the large difference between the effects of $K\bar{K}$ channel in $G_E$ and $G_M$ is the result of the much larger magnitude that $G_M$ has overall than $G_E$. 
The absolute effect of the $K\bar{K}$ inelasticity in $G_M$ is actually of compatible size as in $G_E$ (sometimes even larger).

%
	\begin{figure*}[htbp]
		\centering
		\includegraphics[width=0.45\linewidth,clip]{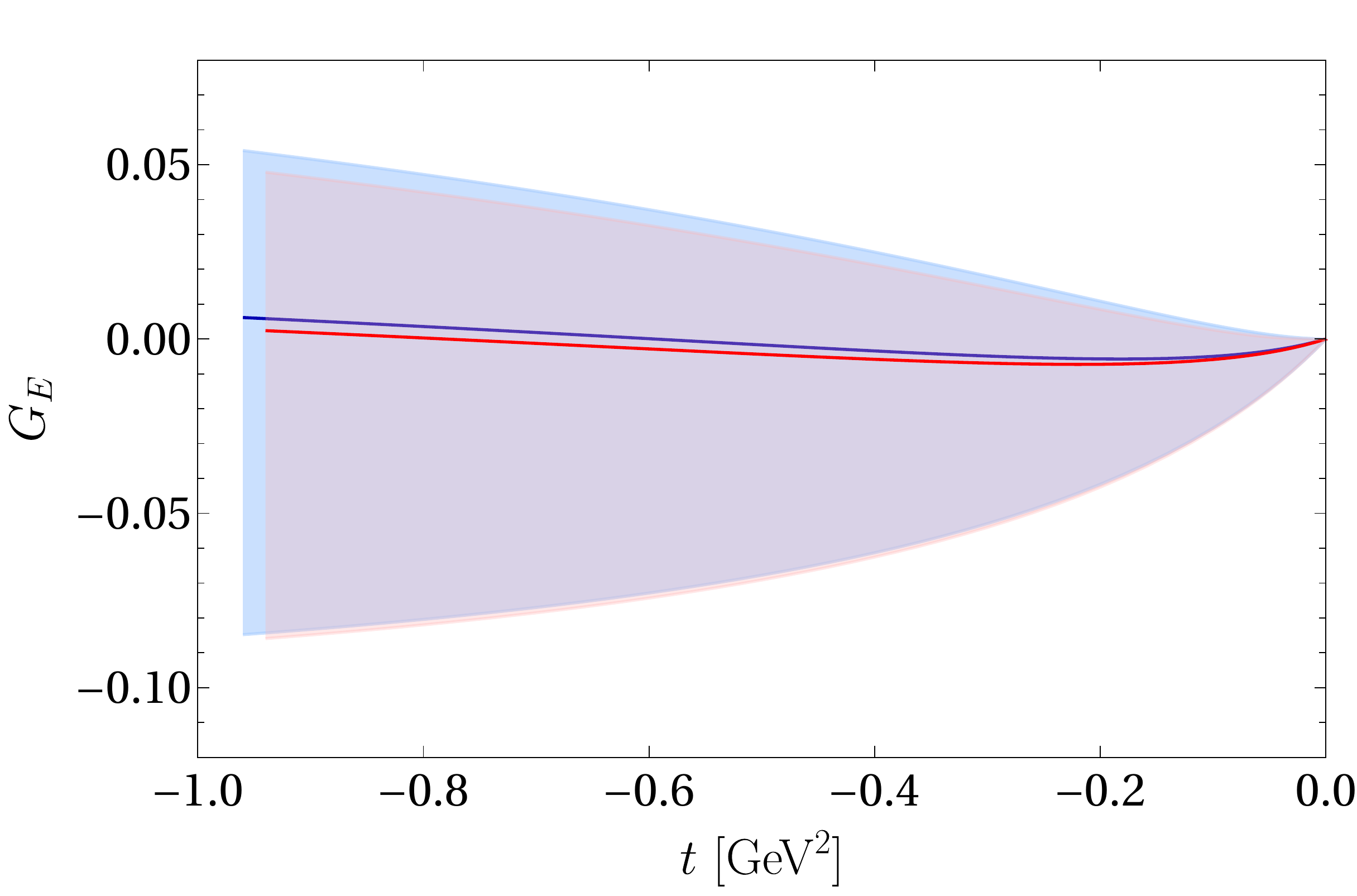}
		\includegraphics[width=0.45\linewidth,clip]{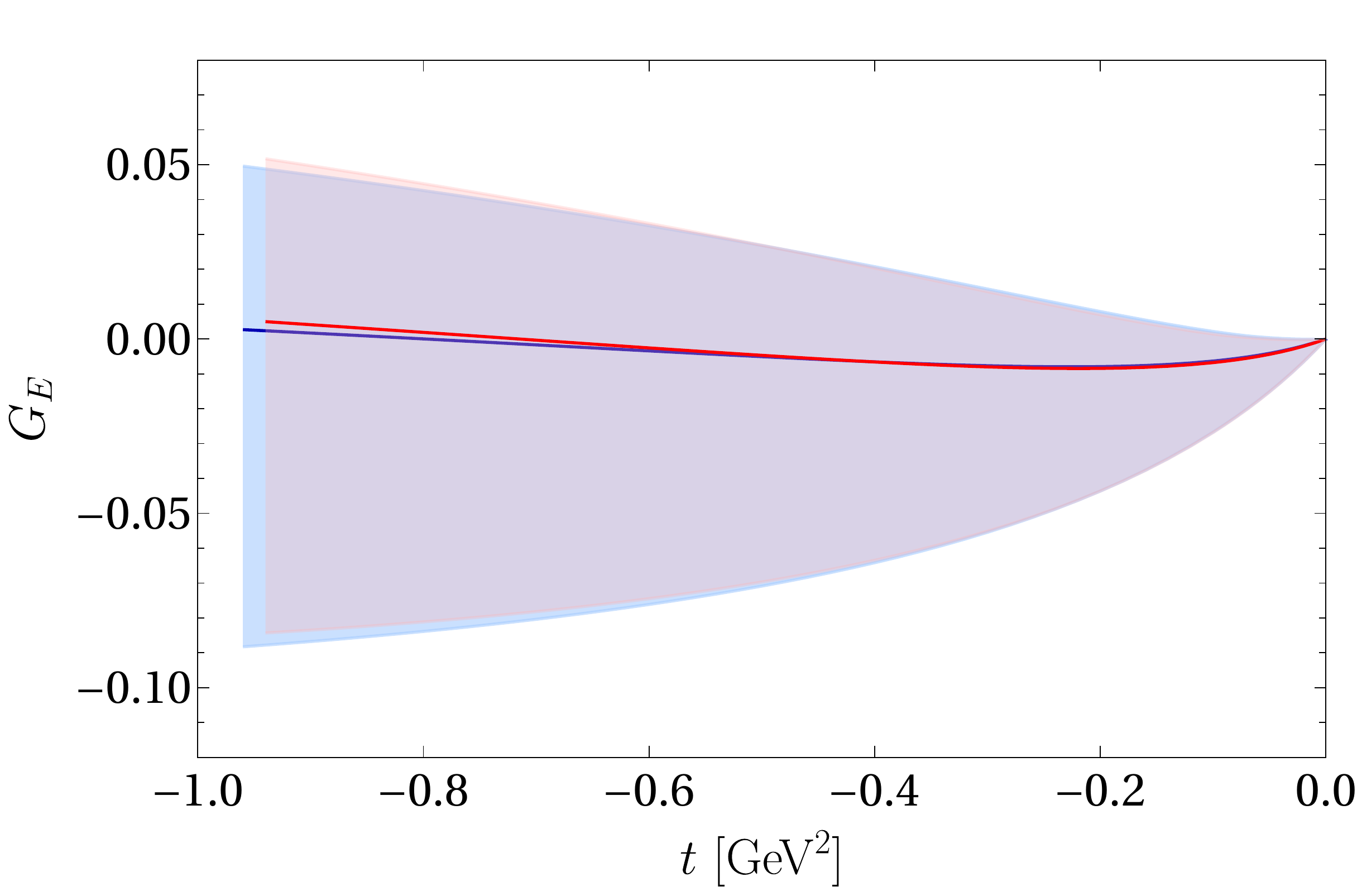}
		\caption{The electric transition form factor $G_E$ obtained from the once-subtracted dispersion 
			relation Eq.~\eqref{eq:tffDR} with an energy cutoff $\Lambda=1.5~\rm GeV$ (left) and $2.0~\rm GeV$ (right).
			The blue lines denote the results from the single $\pi\pi$ channel consideration as in
			Ref.~\cite{Granados:2017cib} and the red lines are those after including the $K\bar{K}$ channel.
			The error bands are estimated based on bootstrap
			sampling. }
		\label{fig:GE}
		\vspace{-3mm}
	\end{figure*}
\begin{figure*}[htbp]
	\centering
	\includegraphics[width=0.45\linewidth,clip]{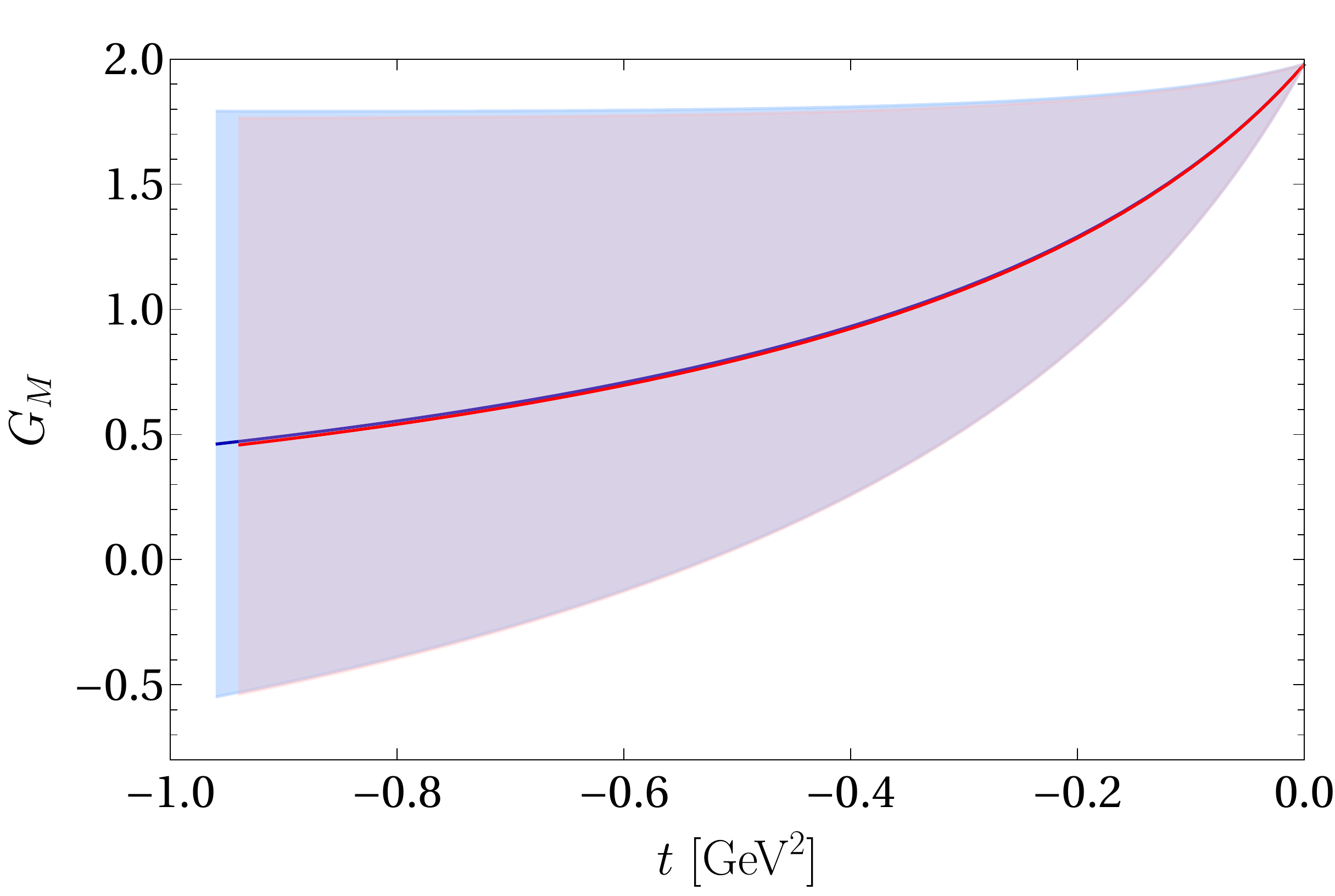}
	\includegraphics[width=0.45\linewidth,clip]{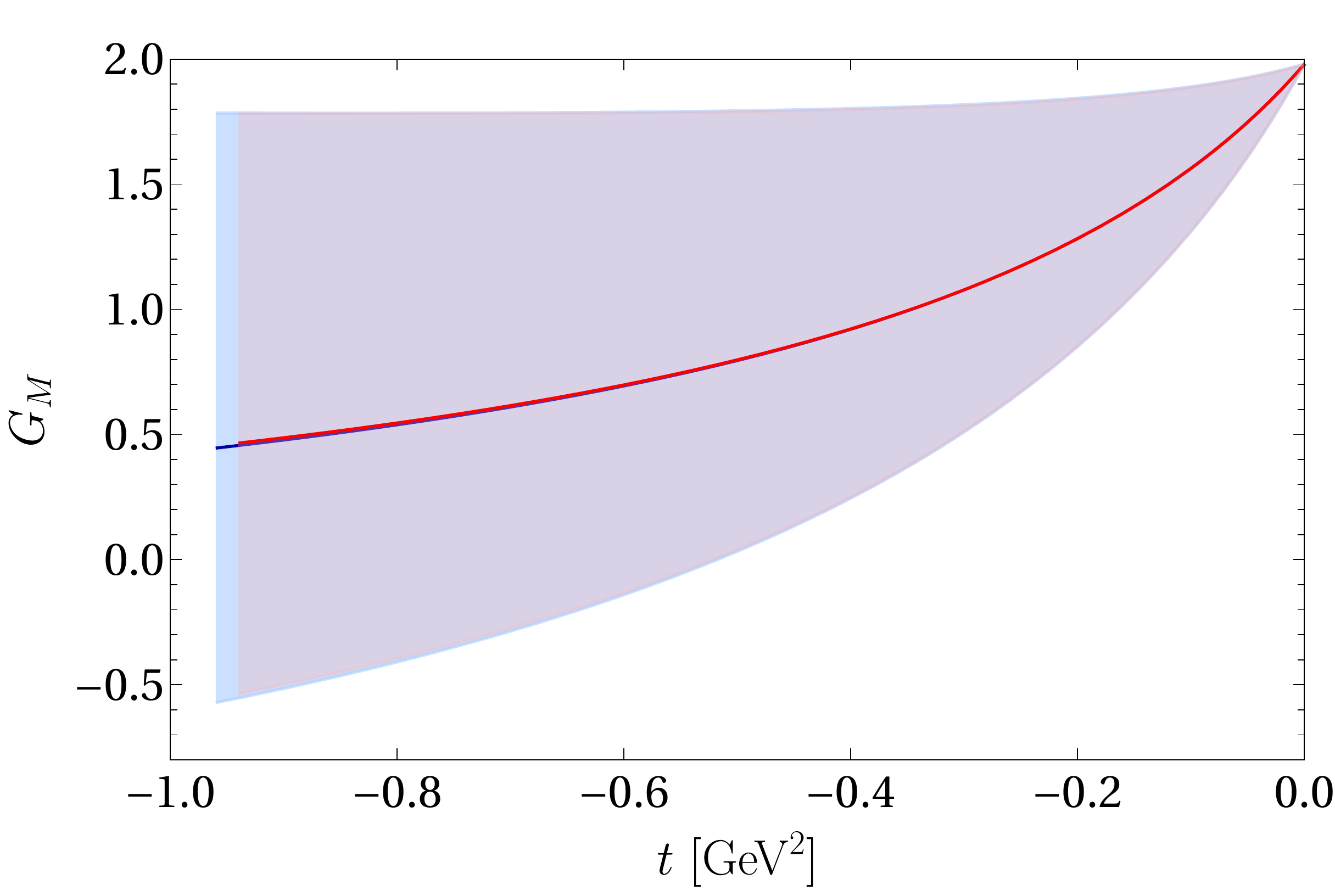}
	\caption{The magnetic transition form factor $G_M$ obtained from the once-subtracted dispersion
		relation Eq.~\eqref{eq:tffDR} with an energy cutoff $\Lambda=1.5~\rm GeV$ (left) and
		$2.0~\rm GeV$ (right). For notations, see Fig.~\ref{fig:GE}. }
	\label{fig:GM}
\end{figure*}
%
In Fig.~\ref{fig:GE}, we show the electric transition
form factor $G_E$ between the estimation including only the $\pi\pi$ intermediate state and
the $\pi\pi$-$K\bar{K}$ coupled-channel determination with errors. Note that the TFFs are real-valued in the space-like region. The solid curves are calculated again with the
radius-adjusted parameters. The error bands in Fig.~\ref{fig:GE} are estimated by the bootstrap sampling over the
three-dimensional parameter space that is spanned by $F_\Phi$, $b_{10}$ and $h_A$. Note that the electric form factor is independent of the low-energy constant $b_{10}$, see the expressions in Appendix~\ref{app:chpt}.
As in Ref.~\cite{Granados:2017cib}, the uncertainty in $h_A$ gives the dominant contribution.
The effect on $G_E$ introduced by the inclusion of the $K\bar{K}$ inelasticity is heavily intertwined with the large uncertainties from the variation of $h_A$ and $\Lambda$. 
Overall, the role of the cutoff is a bit more complicated than in the single $\pi\pi$ channel case.
The situation is different for $G_M$ which is
displayed in Fig.~\ref{fig:GM}. The magnetic Sigma-to-Lambda transition form factor $G_M$
is almost unchanged after including the $K\bar{K}$ inelasticity.
Moreover, $G_M$ has much larger absolute errors from the bootstrap method. At $t=-1$ GeV$^2$, the bootstrap uncertainty from
$F_\Phi$, $h_A$ and $b_{10}$ is already of order $\pm 1$, dominated by the uncertainty in $b_{10}$.
As in Ref.~\cite{Granados:2017cib}, we find a very small sensitivity of $G_M$ to the variation of the
cutoff $\Lambda$.
In addition to providing valuable insights into the electromagnetic structure of hyperons,
experimental data for the transition form factors may thus also help to constrain
these parameters.

\section{Summary}\label{sec:summ}

In this paper, we extended the dispersion theoretical determination of the electromagnetic
Sigma-to-Lambda transition form factors presented in Ref.~\cite{Granados:2017cib} from the
$\pi\pi$ intermediate state to the $\pi\pi$-$K\bar{K}$ coupled-channel configuration within
the SU(3) ChPT framework. After including the $K\bar{K}$ channel, a shift of the electric Sigma-to-Lambda
transition form factor $G_E$ is presented, while the magnetic form factor $G_M$ stays essentially unchanged.
At present, the dispersion theoretical determination of electromagnetic  Sigma-to-Lambda transition form
factors suffers from sizeable uncertainties due to the poor knowledge of the LEC $b_{10}$ and coupling $h_A$. The precise determination of
this three-flavor LEC from the future experiments will be helpful to pin down the hyperon TFFs. In a next step,
it will be of interest to explore the elastic hyperon electromagnetic form factors based on the
theoretical framework that combines dispersion theory and three-flavor chiral perturbation theory.   

\section*{Acknowledgements}
YHL is grateful to Meng-Lin Du, De-Liang Yao and Feng-Kun Guo for many valuable discussions. YHL thanks also Yu-Ji Shi for some discussions on the kaon vector form factors. This work of UGM and YHL is supported in
part by  the DFG (Project number 196253076 - TRR 110)
and the NSFC (Grant No. 11621131001) through the funds provided
to the Sino-German CRC 110 ``Symmetries and the Emergence of
Structure in QCD",  by the Chinese Academy of Sciences (CAS) through a President's
International Fellowship Initiative (PIFI) (Grant No. 2018DM0034), by the VolkswagenStiftung
(Grant No. 93562), and by the EU Horizon 2020 research and innovation programme, STRONG-2020 project
under grant agreement No 824093. HWH was supported by the Deutsche Forschungsgemeinschaft (DFG, German
Research Foundation) -- Projektnummer 279384907 -- CRC 1245
and by the German Federal Ministry of Education and Research (BMBF) (Grant
no. 05P21RDFNB).

\appendix
\section{Unitarity relations and the anomalous pieces}\label{app:app1}
Let us start from the single channel case. The unitarity relations for the $\Sigma$-to-$\Lambda$ TFFs $G_{E/M}$ (in the followings we drop the index $E/M$) within the single $\pi\pi$ channel assumption read~\cite{Granados:2017cib,Junker:2019vvy}
\begin{equation}\label{eq:pipiUni}
	\frac{1}{2 i}{\rm disc}_{\rm unit}~G(t)=\frac{1}{24\pi}T_\pi\Sigma_\pi F_\pi^{V*}~,
\end{equation}
where $\Sigma_\pi=\sigma_\pi q_\pi^2$ with $\sigma$ and $q$ defined by Eq.~\eqref{eq:sigma} and Eq.~\eqref{eq:qcm} respectively, and $q=\sqrt{t}\sigma/2$. Moving to the $\pi\pi$-$K\bar{K}$ coupled-channel case,
one first considers the vector pion and kaon form factors; they satisfy the unitarity relations~\cite{Hoferichter:2012wf,Ditsche:2012fv},
\begin{equation}
	\frac{1}{2i}{\rm disc}\ \vec{F}^V(t)=\vec{t}_1^{1\,*} \vec\Sigma \vec{F}^V,\quad \vec{F}^V=\begin{pmatrix} F_\pi^V, & \sqrt{2}F_K^V\end{pmatrix}^T.
\end{equation}
Similarly, the $\Sigma^0\bar{\Lambda}\to\pi\pi$ and $\Sigma^0\bar{\Lambda}\to K\bar{K}$ $P$-wave amplitudes fulfill the unitarity relations
\begin{equation}\label{eq:t-uni}
	\frac{1}{2i}{\rm disc}\ \vec{T}(t)=\vec{t}_1^{1\,*}\vec\Sigma \vec{T},\quad \vec{T}=\begin{pmatrix}  T_\pi, & \sqrt{2}T_K\end{pmatrix}^T.
\end{equation}
The key information that the above two equations provide us is the relative ratio between the $\pi\pi$ and $K\bar{K}$ channels in the $J=I=1$ coupled-channel problem. Then with the single-$\pi\pi$ unitarity relations at hand already, that is, Eq.~\eqref{eq:pipiUni}, one can easily extend to the two-channel case:
\begin{align}
	\label{eq:twouni}
	&\frac1{2i}{\rm disc}_{\rm unit}~G(t)=\frac{1}{24\pi}\vec{T}^T\vec{\Sigma} \vec{F}^{V*}\notag\\
	&\phantom{mm}=\frac{1}{24\pi}\begin{pmatrix} T_\pi, & \sqrt{2}T_K \end{pmatrix}.\begin{pmatrix} \displaystyle\Sigma_\pi& 0 \\ 0 & \displaystyle\Sigma_K\end{pmatrix}.\begin{pmatrix}  F_\pi^{V*} \\  \sqrt{2}F_\pi^{V*}\end{pmatrix}\notag\\
	&\phantom{mm}=\frac{1}{24\pi}\bigg(T_\pi\Sigma_\pi F_\pi^{V*}\,
	\theta\left(t-4M_\pi^2\right)~\notag\\
	&\phantom{mmmm}+2T_K\Sigma_K F_K^{V*}\, \theta\left(t-4M_K^2\right)\bigg)~.
\end{align}
That becomes Eq.~\eqref{eq:unitarity_relation} after substituting the identity $q=\sqrt{t}\sigma/2$. Recalling that all the left-hand cut (LHC) part of $ T$ is included in $ K$, then ${T}-{K}$ only contains the right-hand cut (RHC) and its unitarity relation is given by Eq.~\eqref{eq:t-uni} for the two-channel assumption. One can also write~\cite{Hoferichter:2012wf}
\begin{equation}\label{eq:Tdisc}
	\frac{1}{2i}{\rm disc}\ \vec{\Omega}^{-1}(\vec{T}-\vec{K})=-\left[{\rm Im}\vec{\Omega}^{-1}\right] \vec{K},
\end{equation}
which leads to Eq.~\eqref{eq:coupled-channel}. 

When $m_\Sigma^2+m_\Lambda^2-2M_{i}^2>2m_{\rm exch}^2$ and $\lambda(m_\Lambda^2,m_{\rm exch}^2,M_i^2)<0$ with $M_i=M_\pi$ ($M_K$) for the process $\Sigma\bar{\Lambda}\to\pi\pi$ ($\Sigma\bar{\Lambda}\to K\bar{K}$), the LHC and RHC will overlap, leading to the non-zero anomalous terms $G_{\rm anom}$ and $\vec{T}_{\rm anom}$ in Eq.~\eqref{eq:unitarity_relation} and Eq.~\eqref{eq:coupled-channel}, respectively~\cite{Karplus:1958zz,Lucha:2006vc,Hoferichter:2013ama,Molnar:2019uos,Junker:2019vvy}. This indeed happens in the proton exchange diagram for the process $\Sigma\bar{\Lambda}\to K\bar{K}$. Such anomalous contributions are estimated by the dispersive integrals of the discontinuity along the cut that connects the anomalous threshold to the starting point of the RHC (the physical threshold of the two-body intermediate state). 
The anomalous threshold $t_-$ is defined by~\cite{Lucha:2006vc}
\begin{align}\label{eq:anompos}
	&t_-=\frac12(m_\Sigma^2+m_\Lambda^2+2M_K^2-m_N^2)\notag\\
	&-\frac1{2m_N^2}\bigg((m_\Sigma^2-M_K^2)(m_\Lambda^2-M_K^2)\notag\\
	&+\lambda^{1/2}(m_\Sigma^2,m_N^2,M_K^2)\lambda^{1/2}(m_\Lambda^2,m_N^2,M_K^2)\bigg).	
\end{align}
Numerically, $t_-=0.935~{\rm GeV}$ located at the real axis of $t$ just below the $K\bar{K}$ threshold. To go further, one first has to derive the discontinuity along the anomalous cut for the TFFs $G$ and the scattering amplitudes $\vec T$. After implementing the partial-wave projection, namely the integration in Eq.~\eqref{eq:kE} and Eq.~\eqref{eq:kM}, one obtains
\begin{equation}
	K_N=\frac{f}{\kappa^3}\log\frac{Y+\kappa}{Y-\kappa}+{\text{remainder}},
\end{equation}
where
\begin{align}\label{eq:Ykappa}
	&Y=-(m_\Sigma^2+m_\Lambda^2+2M_K^2-t-2m_N^2),\notag\\
	&\kappa=\lambda^{1/2}(t,m_\Sigma^2,m_\Lambda^2)\sigma_K(t).
\end{align}
$f$ is the coefficient of the logarithm which is a smooth function over the transferred momentum square $t$ without any cut. The anomalous threshold is generated by the logarithm function. As illustrated in Refs.~\cite{Lucha:2006vc,Junker:2019vvy}, the discontinuity of $K_N$ along the anomalous cut reads
\begin{equation}\label{eq:anomK}
	\frac{1}{2i}{\rm disc}_{\rm anom}\  K_N=\frac{f}{\kappa^2}\frac{2\pi}{(-\lambda(t,m_\Sigma^2,m_\Lambda^2))^{1/2}\sigma_K}.
\end{equation}
Note that the argument of $\sqrt{z}$ is defined in the range of $[0,\pi)$ in the present work. Regarding $\vec T$, one can rewrite Eq.~\eqref{eq:Tdisc} into
\begin{align}
	&\frac{1}{2i}{\rm disc}_{\rm anom}\ \vec{\Omega}^{-1}(\vec{T}-\vec{K})=-\left[{\rm Im}\vec{\Omega}^{-1}\right] \frac{1}{2i}{\rm disc}_{\rm anom}\ \vec{K}\notag\\
	&\phantom{mm}=\left(\vec{\Omega}^{-1}\vec{t}_1^{1*}\vec{\Sigma}\right)\frac{1}{2i}{\rm disc}_{\rm anom}\ \vec{K},\notag
\end{align}
where we replace ($-\left[{\rm Im}\vec{\Omega}^{-1}\right]$) with ($\vec{\Omega}^{-1}\vec{t}_1^{1*}\vec{\Sigma}$) in the second line since $\left[{\rm Im}{\Omega}^{-1}\right]_{12}=\left[{\rm Im}{\Omega}^{-1}\right]_{22}=0$ below the $K\bar{K}$ threshold.\footnote{This replacement is necessary since $\left[{\rm Im}\vec{\Omega}^{-1}\right]$ is solved numerically in our calculation and $\left[{\rm Im}{\Omega}^{-1}\right]_{12}=\left[{\rm Im}{\Omega}^{-1}\right]_{22}=0$ always holds in the unphysical region. The combined quantity $\vec{\Omega}^{-1}\vec{t}_1^{1*}\vec{\Sigma}$ can be simplified analytically when multiplied to ${\rm disc}_{\rm anom}\ \vec{K}$. Then it turns out that the products $\vec{\Omega}^{-1}\vec{t}_1^{1*}$ and $\vec{\Sigma}\, {\rm disc}_{\rm anom}\ \vec{K}$, respectively, are finite along the anomalous cut. Moreover, the identity $-\left[{\rm Im}\vec{\Omega}^{-1}\right]=\vec{\Omega}^{-1}\vec{t}_1^{1*}\vec{\Sigma}$ is checked numerically and does hold near the $K\bar{K}$ threshold.} Finally, the discontinuity of the TFFs $G$ along the anomalous cut can be read off straightforwardly in terms of that of $\vec{T}$,
\begin{align}
	&\frac1{2i}{\rm disc}_{\rm anom}\, G=\frac1{24\pi}\left( (\vec{t}_1^{1*})^{-1} \frac1{2i}{\rm disc}_{\rm anom}\, (\vec T-\vec K)\right)^T \vec{F}^{V*}\notag\\
	&=\frac1{24\pi}\left( (\vec{t}_1^{1*})^{-1} \vec{\Omega}\left(\vec{\Omega}^{-1}\vec{t}_1^{1*}\vec{\Sigma}\right)\frac{1}{2i}{\rm disc}_{\rm anom}\ \vec{K}\right)^T \vec{F}^{V*}\notag\\
	&=\frac1{24\pi}\frac{1}{2i}({\rm disc}_{\rm anom}\ \vec{K})^T\vec\Sigma\vec{F}^{V*}.
\end{align}
Substituting Eq.~\eqref{eq:anomK} into the above equation, one obtains
\begin{equation}
	\frac1{2i}{\rm disc}_{\rm anom}\, G=\frac1{24}\frac{- f F_K^{V*}t}{(-\lambda(t,m_\Sigma^2,m_\Lambda^2))^{3/2}}.
\end{equation}
Then we arrive at the expressions for $G_{\rm anom}$ and $\vec{T}_{\rm anom}$. They are
\begin{align}
	\label{eq:anomG}
	&G_{\rm anom}(t)=\frac{t}{24\pi}\int_{0}^{1} {d x}\frac{d t^\prime(x)}{d x}\frac1{t^\prime(x)-t}\notag\\
	&\phantom{mmm}\times\frac{- f(t^\prime(x)) F_K^{V*}(t^\prime(x))}{(-\lambda(t^\prime(x),m_\Sigma^2,m_\Lambda^2))^{3/2}}~,
\end{align}
\begin{align}
	\label{eq:anomTf}
	&\vec{T}_{\rm anom}(t)=\vec\Omega(t)\frac{t}{\pi}\int_{0}^{1} {d x}\frac{d t^\prime(x)}{d x}\frac1{t^\prime(x)-t}\notag\\
	&\times\frac{\left(\vec{\Omega}^{-1}\vec{t}_1^{1*}\vec{\Sigma}\right)\frac{1}{2i}{\rm disc}_{\rm anom}\ \vec{K}}{t^\prime(t^\prime-t-i\epsilon)}~,
\end{align}
with $t^\prime(x)=(1-x)\ t_-+x\ 4M_K^2$.

\begin{figure}[t]
	\centerline{\includegraphics*[width=0.45\textwidth,angle=0]{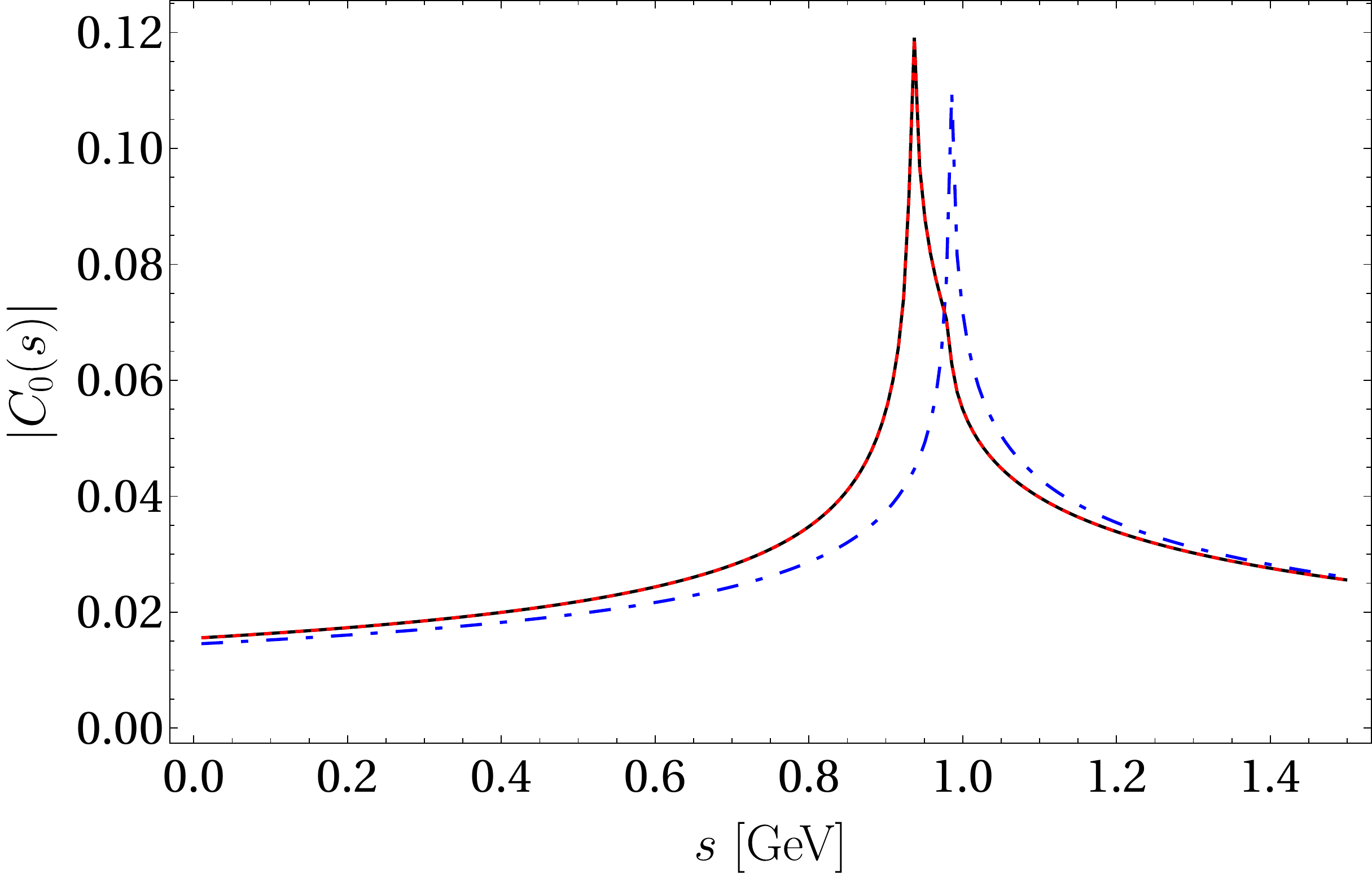}}
	\caption{
		The absolute values of the scalar triangle loop function $C_0(m_\Sigma^2,m_\Lambda^2,
		s,M_K^2,m_N^2,M_K^2)$ calculated numerically using Feynman parameters (solid black line) as well as dispersively with (dashed red line) and without the anomalous contribution (dot-dashed blue line). Note that the solid black and the
                dashed red line coincide.
	}
	\label{fig:triangle}
	\vspace{-3mm}
\end{figure}

To cross-check whether this prescription is correct, we present the calculation of a scalar triangle loop function $C_0(m_\Sigma^2,m_\Lambda^2,s,M_K^2,m_N^2,M_K^2)$ in Fig.~\ref{fig:triangle}. The exact agreement is achieved only when the anomalous contribution is taken into account. 

\section{The reduced amplitudes $K_{\pi/K}$ and $P_{0, \pi/K}$}\label{app:chpt}
\begin{figure}[htbp]
	\centerline{\includegraphics*[width=0.45\textwidth,angle=0]{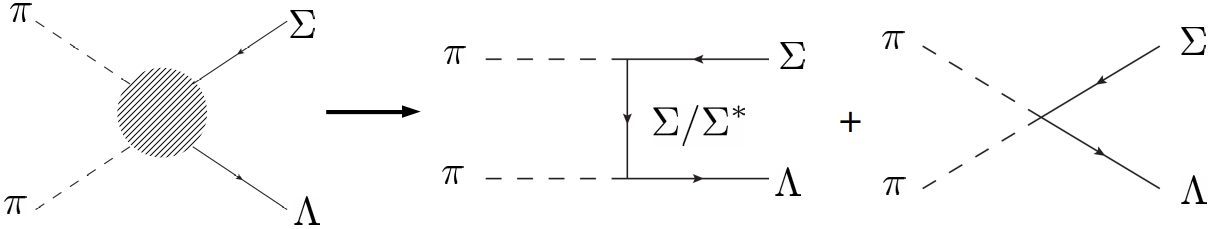}}
	\caption{
		Pictorial representation of the bare input of the four-point amplitude $\pi\pi\to\Sigma^0\bar\Lambda$ obtained up to NLO. 
	}
	\label{fig:Tpipi}
	\vspace{-3mm}
\end{figure}

\begin{figure}[htbp]
	\centerline{\includegraphics*[width=0.45\textwidth,angle=0]{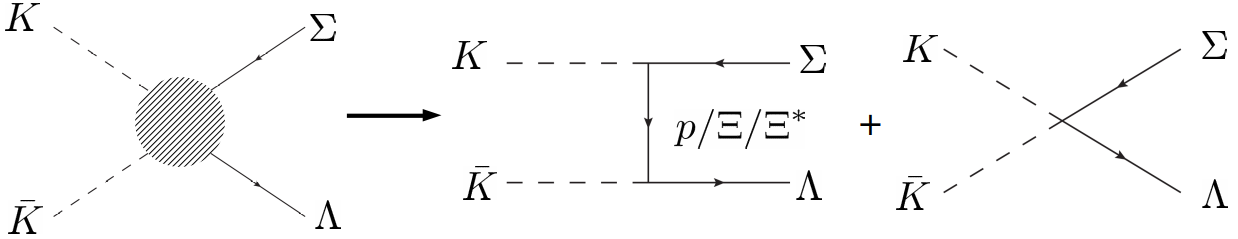}}
	\caption{
		Pictorial representation of the bare input of the four-point amplitude $K\bar{K}\to\Sigma^0\bar\Lambda$ obtained up to NLO.
	}
	\label{fig:Tkk}
	\vspace{-3mm}
\end{figure}

The four-point amplitudes ${\cal M}_{\Sigma\bar\Lambda\to\pi\pi/{K\bar{K}}}(t,\theta)$ are calculated up to next-to leading order within the framework of  SU(3) chiral perturbation theory. It turns out that the explicit inclusion of the decuplet baryon in the three-flavor ChPT Lagrangian is important to reproduce the correct $G_{E/M}(0)$~\footnote{Here, the normalization of electromagnetic Sigma-to-Lambda TFFs is estimated with the unsubtract dispersion relations, see Ref.~\cite{Granados:2017cib} for more details. } and reasonable electric and magnetic transition radii, $\langle r_E^2\rangle$ and $\langle r_M^2\rangle$~\cite{Granados:2017cib}. We use the same Lagragians as in Ref.~\cite{Granados:2017cib}. To be specific, the relevant interaction part of the leading order (LO) chiral Lagrangian that contains both the octet and decuplet states as active degrees of freedom for the reactions of interest is given by~\cite{Kubis:2000aa,Ledwig:2014rfa}
\begin{align}\label{eq:lagrangian}
	{\cal L}_{8+10}^{(1)}&=\frac{D}2\langle \bar{B}\gamma^\mu\gamma_5\{u_\mu,B\}\rangle+\frac{F}2\langle \bar{B}\gamma^\mu\gamma_5[u_\mu,B]\rangle\notag\\
	&+\frac1{2\sqrt{2}}h_A\epsilon_{ade}g_{\mu\nu}(\bar{T}_{abc}^\mu u_{bd}^\nu B_{ce}+\bar{B}_{ec} u_{db}^\nu T_{abc}^\mu),
\end{align}
and the relevant NLO Lagrangian reads~\cite{Oller:2006yh,Frink:2006hx}
\begin{equation}
	{\cal L}_8^{(2)}=\frac{i}2 b_{10}\langle\bar{B}\{[u^\mu,u^\nu],\sigma_{\mu\nu}B\}\rangle,
\end{equation}
where $\langle\cdots\rangle$ denotes a flavor trace. The chirally covariant derivatives are defined by
\begin{eqnarray}
	\label{eq:devder}
	D^\mu B := \partial^\mu B + [\Gamma^\mu,B]
\end{eqnarray}
with
\begin{eqnarray}
	\Gamma_\mu &= & \frac12 \, \left(
	u^\dagger \, \left( \partial_\mu - i (v_\mu + a_\mu) \right) \, u \right. \nonumber \\ && \left. {}+
	u \, \left( \partial_\mu - i (v_\mu - a_\mu) \right) \, u^\dagger
	\right) \,.
	\label{eq:defGammamu}
\end{eqnarray}
Here, $v$ and $a$ are external sources and $u^2=U=\exp(i\Phi/F_\Phi)$ with the Goldstone bosons encoded in
the matrix
\begin{eqnarray}
	\Phi &=&  \left(
	\begin{array}{ccc}
		\pi^0 +\frac{1}{\sqrt{3}}\, \eta 
		& \sqrt{2}\, \pi^+ & \sqrt{2} \, K^+ \\
		\sqrt{2}\, \pi^- & -\pi^0+\frac{1}{\sqrt{3}}\, \eta
		& \sqrt{2} \, K^0 \\
		\sqrt{2}\, K^- & \sqrt{2} \, {\bar{K}}^0 
		& -\frac{2}{\sqrt{3}}\, \eta
	\end{array}   
	\right) 
	\,. \label{eq:gold1}
\end{eqnarray}
The octet baryons also make up a $3\times3$ matrix in the flavor space that is given by
\begin{eqnarray}
	\label{eq:baroct}
	B = \left(
	\begin{array}{ccc}
		\frac{1}{\sqrt{2}}\, \Sigma^0 +\frac{1}{\sqrt{6}}\, \Lambda 
		& \Sigma^+ & p \\
		\Sigma^- & -\frac{1}{\sqrt{2}}\,\Sigma^0+\frac{1}{\sqrt{6}}\, \Lambda
		& n \\
		\Xi^- & \Xi^0 
		& -\frac{2}{\sqrt{6}}\, \Lambda
	\end{array}   
	\right)  \,.
\end{eqnarray}
Finally, $T_{abc}$ is a totally symmetric flavor tensor that denotes the decuplet baryons,
\begin{eqnarray}
	&& T_{111} = \Delta^{++} \,, \quad T_{112} = \frac{1}{\sqrt{3}} \, \Delta^+  \,, \nonumber \\
	&& T_{122} = \frac{1}{\sqrt{3}} \, \Delta^0  \,,  T_{222} = \Delta^- \,, \nonumber \\
	&& T_{113} = \frac{1}{\sqrt{3}} \, \Sigma^{*+}  \,, \quad T_{123} = \frac{1}{\sqrt{6}} \, \Sigma^{*0}  \,, \quad 
	T_{223} = \frac{1}{\sqrt{3}} \, \Sigma^{*-}  \,,  \nonumber \\
	&& T_{133} = \frac{1}{\sqrt{3}} \, \Xi^{*0} \,, \quad T_{233} = \frac{1}{\sqrt{3}} \, \Xi^{*-} \,, \quad 
	T_{333} = \Omega \,. 
	\label{eq:tensorT}
\end{eqnarray}
The amplitudes ${\cal M}_{\Sigma\bar\Lambda\to\pi\pi/{K\bar{K}}}$ are described as a
Born term in the LO plus a contact term in the NLO within the three-flavor ChPT, see Fig.~\ref{fig:Tpipi} and Fig.~\ref{fig:Tkk}.

From above Lagrangians, one obtains the $\Sigma$-exchange Born term for $\Sigma^0(p_1)+\bar{\Lambda}(p_2)\to\pi^-(p_3)+\pi^+(p_4)$,

\begin{align}
	&\phantom{mmmm}i {\cal M}_{\Sigma} =i( {\cal M}_{\rm t}+{\cal M}_{\rm u})\notag\\
	&i{\cal M}_{\rm t}=  \frac{D F}{\sqrt{3}F_\Phi^2}\Bigg(\bar v_\Lambda \gamma^\mu \gamma_5 p_{3,\mu}S_{\Sigma^-,t}\gamma^\nu\gamma_5
	p_{4,\nu} u_\Sigma 
	\Bigg),\notag\\
	&i{\cal M}_{\rm u}=  \frac{-D F}{\sqrt{3}F_\Phi^2}\Bigg(\bar v_\Lambda \gamma^\mu \gamma_5 p_{4,\mu}S_{\Sigma^+,u}\gamma^\nu\gamma_5
	p_{3,\nu} u_\Sigma 
	\Bigg),
	\label{eq:bnsigma}
\end{align}
with $S_{\Sigma^-,t}=i((p_1-p_4)^\mu\gamma_\mu+m_\Sigma)/(t-m_\Sigma^2)$ and $S_{\Sigma^+,u}=i((p_1-p_3)^\mu\gamma_\mu+m_\Sigma)/(u-m_\Sigma^2)$ the propagator of the exchanged $\Sigma$ in the $t$- and $u$-channel respectively. And the $\Sigma^*$-exchange Born term,
\begin{align}
	&i {\cal M}_{\Sigma^*} =i( {\cal M}_{\rm t}+{\cal M}_{\rm u})\notag\\
	&=\left(\frac{-h_A}{2\sqrt{2}F_\Phi}\right)^2\bar{v}_\Lambda g_{\mu\nu}p_3^\nu \Delta^{\mu\alpha}_t(\frac{-1}{\sqrt{3}})g_{\alpha\beta}p_4^\beta u_\Sigma,\notag\\
	&+\left(\frac{-h_A}{2\sqrt{2}F_\Phi}\right)^2\bar{v}_\Lambda (-1) g_{\mu\nu}p_4^\nu \Delta^{\mu\alpha}_u(\frac{-1}{\sqrt{3}})g_{\alpha\beta}p_3^\beta u_\Sigma,
	\label{eq:bnsgmastar}
\end{align}
with the spin-$3/2$ Rarita-Schwinger propagator~\cite{Pascalutsa:1999zz}
\begin{align}
	i\Delta^{\mu\nu}(p)&=\frac{\gamma^\alpha p_\alpha+m}{p^2-m^2}\bigg(g^{\mu\nu}-\frac13 \gamma^\mu\gamma^\nu\notag\\
	&-\frac1{3p^2}\gamma_\alpha\gamma_\beta p_\rho p_\lambda( g^{\mu\beta }g^{\nu\lambda} g^{\alpha\rho}+g^{\nu\alpha }g^{\mu\rho} g^{\beta\lambda})\bigg)\notag\\
	&-\frac2{3\, m^2}\frac{p^\mu p^\nu}{p^2}(\gamma^\alpha p_\alpha+m)\notag\\
	&+\frac{-i}{3\, m\, p^2}(g^{\mu\rho}g^{\nu\beta}g^{\alpha\lambda}+g^{\mu\alpha}g^{\nu\lambda}g^{\beta\rho})\sigma_{\alpha\beta}p_\rho p_\lambda\notag,
\end{align}
and $t=(p_1-p_4)^2$, $u=(p_1-p_3)^2$. Here, $m$ denotes the mass of the exchanged spin-$3/2$ resonance. The NLO contact term for the reaction $\Sigma^0(p_1)+\bar{\Lambda}(p_2)\to\pi^-(p_3)+\pi^+(p_4)$ is given by
\begin{align}
	&{\cal M}_{\rm NLO}=\left(b_{10}\frac1{F_\Phi^2}\frac4{\sqrt{3}}\right)\frac12\,\times\notag\\
	&\bigg((m_\Sigma+m_\Lambda)\left(-\bar{v}_\Lambda\gamma^\mu( p_4- p_3 )_\mu u_\Sigma\right)+(u-t)\bar{v}_\Lambda u_\Sigma
	\bigg)~.
	\label{eq:NLOpi}
\end{align}
The corresponding expressions for the $\Sigma^0(p_1)+\bar{\Lambda}(p_2)\to K^-(p_3)+K^+(p_4)$ (${\cal M}_{\Sigma^0\bar\Lambda\to{K^0\bar{K}^0}}=-{\cal M}_{\Sigma^0\bar\Lambda\to{K^+{K}^-}}$ in the isospin limit) read
\begin{align}
	&i {\cal M}_{\rm born}=i( {\cal M}_{\rm u}+{\cal M}_{\rm t}+{\cal M}_{\Xi^*})~,\notag\\
	&i{\cal M}_{\rm u}=  \frac{1}{ F_\Phi^2} \left(\frac{-D}{2\sqrt{3}}+\frac{-\sqrt{3}F}{2}\right)\frac{D-F}2\,\notag\\
	&\phantom{mmmmmm}\times\Bigg(
	\bar v_\Lambda \gamma^\mu\gamma_5 p_{4,\mu} S_{p,u} \gamma^\nu \gamma_5 p_{3,\nu} u_\Sigma\Bigg)~,\notag
\end{align}
\begin{align}
	&i{\cal M}_{\rm t}=\frac{1}{ F_\Phi^2} \left(\frac{-D}{2\sqrt{3}}+\frac{\sqrt{3}F}{2}\right)\frac{D+F}2 \notag\\
	&\phantom{mmmmmm}\times\Bigg(
	\bar v_\Lambda \gamma^\mu\gamma_5 p_{3,\mu} S_{\Xi,t} \gamma^\nu \gamma_5 p_{4,\nu} u_\Sigma\Bigg)~,\notag\\
	&i{\cal M}_{\Xi^*}=\left(\frac{-h_A}{2\sqrt{2}F_\Phi}\right)^2\bar{v}_\Lambda (+1) g_{\mu\nu}p_3^\nu \Delta^{\mu\alpha}_t(\frac{-1}{\sqrt{3}})g_{\alpha\beta}p_4^\beta u_\Sigma~.\notag\\
	&\phantom{m}{\cal M}_{\rm NLO}=\left(b_{10}\frac1{F_\Phi^2}\frac2{\sqrt{3}}\right)\frac12 \notag\\
	&\times\bigg((m_\Sigma+m_\Lambda)\left(-\bar{v}_\Lambda\gamma^\mu( p_4- p_3 )_\mu u_\Sigma\right)+(u-t)\bar{v}_\Lambda u_\Sigma\bigg)~,
	\label{eq:ampK}
\end{align}
with $S_{p,u}=i((p_1-p_3)^\mu\gamma_\mu+m_p)/(u-m_p^2)$ and $S_{\Xi,t}=i((p_1-p_4)^\mu\gamma_\mu+m_\Xi)/(t-m_\Xi^2)$ the propagator of the exchanged proton and $\Xi$ baryon, respectively. To proceed, it is helpful to introduce the following equivalents,
\begin{align}
	E1&\equiv\frac{\bar{v}_{1/2,\Lambda}\gamma^\mu(p_1-p_2)_\mu u_{1/2,\Sigma}}{\bar{v}_{1/2,\Lambda}\gamma^3u_{1/2,\Sigma}}=\frac{\bar{v}_{1/2,\Lambda}u_{1/2,\Sigma}(m_\Lambda+m_\Sigma)}{\bar{v}_{1/2,\Lambda}\gamma^3u_{1/2,\Sigma}}\notag\\
	&=\frac{(m_\Sigma+m_\Lambda)^2-s}{2p_z}~,\notag\\
	E2&\equiv\frac{\bar{v}_{1/2,\Lambda}\gamma^\mu({p_4-p_3})_\mu u_{1/2,\Sigma}}{\bar{v}_{1/2,\Lambda}\gamma^3u_{1/2,\Sigma}}=-2p_{\rm c.m.}\cos\theta~,\notag\\
	M1&\equiv\frac{\bar{v}_{-1/2,\Lambda}u_{1/2,\Sigma}(m_\Lambda+m_\Sigma)}{\bar{v}_{-1/2,\Lambda}\gamma^1u_{1/2,\Sigma}}=0~,\notag\\
	M2&\equiv\frac{\bar{v}_{-1/2,\Lambda}\gamma^\mu({p_4-p_3})_\mu u_{1/2,\Sigma}}{\bar{v}_{-1/2,\Lambda}\gamma^1u_{1/2,\Sigma}}=-2p_{\rm c.m.}\sin\theta~,
	\label{eq:equavs}
\end{align}
where $s=(p_1+p_2)^2=(p_3+p_4)^2$ is the center-of-mass energy. $p_z$ and $p_{\rm c.m.}$ denote the modulus of the three-dimensional center-of-mass momenta of the $\Sigma\bar\Lambda$ and $\pi\pi/K\bar{K}$ two-body systems, respectively, i.e. $p_{\rm c.m.}=q_{\pi/K}$. The equations~\eqref{eq:equavs} are calculated in the center-of-mass frame with the $p_z$ the modulus of the three-momentum along the direction of the $z$-axis and $\theta$ is the scattering angle of $\pi$ or $K$. Substitute Eqs.~\eqref{eq:bnsigma}, \eqref{eq:bnsgmastar}, \eqref{eq:NLOpi}, \eqref{eq:ampK} into Eqs.~\eqref{eq:kE},\eqref{eq:pE},\eqref{eq:kM}, \eqref{eq:pM}, we obtain $P_{0,\pi}^E$, $P_{0,\pi}^M$, $K_{\pi}^E$ and $K_{\pi}^M$ for the $\pi\pi$ inelasticity,
\begin{align}
	P_{0,\pi}^E  &=  P^E_{\Sigma} + P^E_{\Sigma^*}~,\\
	P^E_{\Sigma}&=\frac32 \, \int\limits_0^\pi d\theta  \sin\theta \,\cos\theta \frac{DF}{\sqrt{3}F_\Phi^2}\frac{E2}{p_{\rm c.m.}}=-\frac2{\sqrt{3}} \frac{DF}{F_\Phi^2}~,\notag\\
	P^E_{\Sigma^*}&=\frac32 \, \int\limits_0^\pi d\theta  \sin\theta \,\cos\theta \left(\frac{-h_A}{2\sqrt{2}F_\pi}\right)^2\frac{1}{\sqrt{3}}\bigg({\frac{t-u}{12 m_{\Sigma^*}^2}\frac{E1}{p_{\rm c.m.}}}\notag\\
	&+\frac{E2}{p_{\rm c.m.}}\frac1{12 m_{\Sigma^*}^2}(-2m_{\Sigma^*}^2-2m_{\Sigma^*}(m_\Sigma+m_\Lambda)+m_\Sigma^2\notag\\
		&\phantom{mmmmmmmmm}+m_\Lambda^2+s-6M_\pi^2)\bigg)
	\notag\\
	&=\frac{h_A^2}{24\sqrt{3}F_\Phi^2}\frac{(m_\Lambda+m_{\Sigma^*})(m_\Sigma+m_{\Sigma^*})}{m_{\Sigma^*}^2}+{\cal O}(M_\pi^2,s)~.\notag
\end{align}
\begin{align}
	K_{\pi}^E  &=  K^E_{\Sigma} + K^E_{\Sigma^*}~,\\
	K^E_{\Sigma}&=\frac32 \, \int\limits_0^\pi d\theta  \sin\theta \,\cos\theta \frac{DF}{\sqrt{3}F_\Phi^2}\notag\\
	&\times\bigg(\frac{E1}{p_{\rm c.m.}}\, m_\Sigma \, (m_\Sigma-m_\Lambda) 
	\left( \frac{1}{t-m_\Sigma^2}-\frac{1}{u-m_\Sigma^2} \right) \notag\\
	& +
	\frac{E2}{p_{\rm c.m.}} \, m_\Sigma \, (m_\Sigma+m_\Lambda) \, 
	\left( \frac{1}{t-m_\Sigma^2}+\frac{1}{u-m_\Sigma^2} \right) \bigg)~,
	\notag\\
	K^E_{\Sigma^*}&=\frac32 \, \int\limits_0^\pi d\theta  \sin\theta \,\cos\theta \left(\frac{-h_A}{2\sqrt{2}F_\pi}\right)^2\frac{1}{\sqrt{3}}\notag\\
	&\times\bigg(
	{+\frac{F(s)}{m_\Sigma+m_\Lambda}\frac{E1}{p_{\rm c.m.}}\left(\frac1{u-m_{\Sigma^*}^2}-\frac1{t-m_{\Sigma^*}^2}\right)}\notag\\
	&{+\frac{E2}{p_{\rm c.m.}}\left(\frac1{u-m_{\Sigma^*}^2}+\frac1{t-m_{\Sigma^*}^2}\right)\frac{G(s)}2}  \bigg)~,\notag
\end{align}
where 
\begin{align}
	& F(s) = \left( \frac{m_\Sigma+m_\Lambda}{2}+m_{\Sigma^*} \right) H_1(s) \notag\\
	&\phantom{mmmmmmmmm}+ \left( \frac{m_\Sigma+m_\Lambda}{2}-m_{\Sigma^*} \right) H_2  \,, \notag \\
	&G(s)  =  H_1(s)+H_2   \,,  \notag \\
	&H_1(s) =  \frac{m_\Sigma^2 + m_\Lambda^2-s}{2}\notag\\
	&\phantom{mmmm}-\frac{(m_\Lambda^2+m_{\Sigma^*}^2-M_\pi^2)(m_\Sigma^2+m_{\Sigma^*}^2-M_\pi^2)}{4 m_{\Sigma^*}^2} \,,  \nonumber  \\
	&H_2 = \frac13 \, \left( m_\Lambda + \frac{m_\Lambda^2+m_{\Sigma^*}^2-M_\pi^2}{2 m_{\Sigma^*}} \right)\notag\\
	&\phantom{mmmmmm}\times \left( m_\Sigma + \frac{m_\Sigma^2+m_{\Sigma^*}^2-M_\pi^2}{2 m_{\Sigma^*}} \right)  \,. \notag 
\end{align}
\begin{align}
	P_{0,\pi}^M  &=  P^M_{\Sigma} + P^M_{\rm NLO}-K^M_{\Sigma^*,\rm low}~,\\
	P^M_{\Sigma}&=\frac34 \, \int\limits_0^\pi d\theta  \sin\theta \,\sin\theta \frac{DF}{\sqrt{3}F_\Phi^2}\frac{M2}{p_{\rm c.m.}}=-\frac2{\sqrt{3}}\frac{DF}{F_\Phi^2}~,\notag\\
	P^M_{\rm NLO}&=\frac34 \, \int\limits_0^\pi d\theta  \sin\theta \,\sin\theta \left(b_{10}\frac1{F_\Phi^2}\frac4{\sqrt{3}}\right)\frac{(m_\Sigma+m_\Lambda)}{2}\frac{-M2}{p_{\rm c.m.}}
	\notag\\
	&=\frac{4}{\sqrt{3}}\frac{b_{10}}{F_\Phi^2}(m_\Lambda+m_\Sigma)~.\notag
\end{align} 
\begin{align}
	K_{\pi}^M  &=  K^M_{\Sigma} + K^M_{\Sigma^*}~,\\
	K^M_{\Sigma}&=\frac34 \, \int\limits_0^\pi d\theta  \sin\theta \,\sin\theta \frac{DF}{\sqrt{3}F_\Phi^2} \frac{M2}{p_{\rm c.m.}}\notag\\
	&\phantom{mmm}\times m_\Sigma \, (m_\Sigma+m_\Lambda) \, 
	\left( \frac{1}{t-m_\Sigma^2}+\frac{1}{u-m_\Sigma^2} \right)~, \notag\\
	K^M_{\Sigma^*}&=\frac34 \, \int\limits_0^\pi d\theta  \sin\theta \,\sin\theta\left(\frac{-h_A}{2\sqrt{2}F_\pi}\right)^2\frac{1}{\sqrt{3}}\notag\\
	&\phantom{mmm}\times\bigg(+\frac{M2}{p_{\rm c.m.}}\left(\frac1{u-m_{\Sigma^*}^2}+\frac1{t-m_{\Sigma^*}^2}\right)\frac{G(s)}2\bigg)~.\notag
\end{align}
Note that we subtract a term $K^M_{\Sigma^*,\rm low}$ in the polynomial part of the magnetic amplitude $P_{0,\pi}^M$, which denotes the low-energy limit of the LHC contribution of the decuplet-exchanged magnetic amplitude. It is proposed to remove the doubly counted decuplet baryon contribution caused by the using of the resonance saturation assumption for the estimation of $b_{10}$ in the present ChPT framework. A similar term $K^E_{\Sigma^*,\rm low}$ should be subtracted in $P_{0,\pi}^E$. However, it belongs to a higher chiral order and is dropped here. Note that $P^E_{\rm NLO}$ belongs to $P_{1}(s)$ that is beyond the accuracy of Eq.~\eqref{eq:coupled-channel} and is also dropped. Taking the same convention with Ref.~\cite{Granados:2017cib}, $K^M_{\Sigma^*,\rm low}$ is given by
\begin{align}
	& K^M_{\Sigma^*,\rm low} = \lim_{s \to 0} \, \lim_{m_\Lambda \to m_\Sigma} \, \lim_{M_\pi \to 0} K^M_{\Sigma^*}(s) \notag\\
	&= \frac{h_A^2}{24 \sqrt{3} F_\Phi^2} \, 
	\frac{(-m_{\Sigma^*}^2+4 m_{\Sigma^*} m_\Sigma - m_\Sigma^2) \, (m_{\Sigma^*} + m_\Sigma)}{m_{\Sigma^*}^2 \, (m_{\Sigma^*} - m_\Sigma)} \,.
	\nonumber 
\end{align}
And similarly, the $P_{0,K}^E$, $P_{0,K}^M$, $K_{K}^E$ and $K_{K}^M$ for the $K\bar{K}$ inelasticity read
\begin{equation}
	P_{0,K}^E  =  P^E_{\rm born} + P^E_{\Xi^*},
\end{equation}
with
\begin{align}
	P^E_{\rm born}&=\frac32 \, \int\limits_0^\pi d\theta  \sin\theta \,\cos\theta \Bigg(\frac12\bigg(g_A(m_\Lambda+m_\Sigma+2m_N)\notag\\
	&\phantom{m}+g_B(m_\Lambda+m_\Sigma+2m_\Xi)\bigg)\frac{E1}{(m_\Lambda+m_\Sigma)p_{\rm c.m.}}\notag\\
	&\phantom{m}+\frac{g_B-g_A}2\frac{E2}{p_{\rm c.m.}}\Bigg)\notag\\
	&=g_A-g_B~,\notag
\end{align}
\begin{align}
	&P^E_{\Xi^*}=\frac32 \, \int\limits_0^\pi d\theta  \sin\theta \,\cos\theta \frac{h_A^2}{8\sqrt{3}F_\Phi^2}\Bigg(
	\bigg(\frac1{{12m_{\Xi^*}^2}}\notag\\
	&\times{(m_\Lambda+m_\Sigma)(t-m_{\Xi^*}^2)}+\frac1{12m_{\Xi^*}^2}(m_\Lambda+m_\Sigma+2m_{\Xi^*})\notag\\
	&\times(-m_\Lambda^2-m_\Sigma^2+2M_K^2+2m_{\Xi^*}^2\notag\\
	&\phantom{mmmmmmmm}+m_{\Xi^*}(m_\Lambda+m_\Sigma))\bigg)\frac{E1}{(m_\Lambda+m_\Sigma)p_{\rm c.m.}}\notag\\
	&+\bigg(\frac1{12}(1-\frac t{m_{\Xi^*}^2})+\frac1{12m_{\Xi^*}^2}(m_\Lambda^2+m_\Sigma^2-2M_K^2\notag\\
	&\phantom{mm}-2m_{\Xi^*}^2-m_{\Xi^*}(m_\Lambda+m_\Sigma))\bigg)\frac{E2}{p_{\rm c.m.}}\Bigg)\notag\\
	&=\frac{h_A^2(m_\Lambda+m_{\Xi^*})(m_\Sigma+m_{\Xi^*})}{48\sqrt{3}F_\Phi^2m_{\Xi^*}^2}+{\cal O}(s,M_K^2)~.\notag
\end{align}
\begin{align}
	K_{K}^E  &=  K^E_{N}+K^E_{\Xi} + K^E_{\Xi^*}~,\\
	K^E_{ N}&=\frac32 \, \int\limits_0^\pi d\theta  \sin\theta \,\cos\theta 
	\bigg(\frac{E1}{(m_\Lambda+m_\Sigma)p_{\rm c.m.}}\notag\\
	&\times g_A\frac{(m_\Lambda+m_N)(m_\Sigma+m_N)(m_\Lambda+m_\Sigma-2m_N)}{2(m_N^2-u)}\notag\\
	&+\frac{E2}{p_{\rm c.m.}}\frac12\frac{g_A(m_\Lambda+m_N)(m_\Sigma+m_N)}{m_N^2-u}\bigg)~,
	\notag\\
	K^E_{\Xi}&=\frac32 \, \int\limits_0^\pi d\theta  \sin\theta \,\cos\theta 
	\bigg(\frac{E1}{(m_\Lambda+m_\Sigma)p_{\rm c.m.}}\notag\\
	&\times g_B\frac{(m_\Lambda+m_\Xi)(m_\Sigma+m_\Xi)(m_\Lambda+m_\Sigma-2m_\Xi)}{2(m_\Xi^2-t)}\notag\\
	&+\frac{E2}{p_{\rm c.m.}}\frac12\bigg(-\frac{g_B(m_\Lambda+m_\Xi)(m_\Sigma+m_\Xi)}{m_\Xi^2-t}\bigg)\bigg)~,
	\notag\\
	K^E_{\Xi^*}&=\frac32 \, \int\limits_0^\pi d\theta  \sin\theta \,\cos\theta \frac{h_A^2}{8\sqrt{3}F_\Phi^2}\Bigg({\frac{-\tilde{F}(s)}{12m_{\Xi^*}^2(m_{\Xi^*}^2-t)}}\notag\\
	&\phantom{m}\times\frac{E1}{(m_\Lambda+m_\Sigma)p_{\rm c.m.}}{+\frac{E2}{p_{\rm c.m.}}\frac{1}{12m_{\Xi^*}^2(m_{\Xi^*}^2-t)}\tilde{G}(s)}\Bigg)~,
	\notag
\end{align}
where 
\begin{align}
	& \tilde{F}(s) = -M_K^4(m_\Lambda+m_\Sigma+4m_{\Xi^*})\notag\\
	&\phantom{mm}+M_K^2(m_\Lambda^3+8m_{\Xi^*}^3+4m_{\Xi^*}^2m_\Sigma+3m_{\Xi^*}m_\Sigma^2+m_\Sigma^3\notag\\
	&\phantom{mm}+m_\Lambda^2(3m_{\Xi^*}+m_\Sigma)+m_\Lambda(4m_{\Xi^*}^2-2m_{\Xi^*}m_\Sigma+m_\Sigma^2))\notag\\
	&\phantom{mm}+(m_\Lambda+m_{\Xi^*})(m_\Sigma+m_{\Xi^*})(m_\Lambda^2(2m_{\Xi^*}-m_\Sigma)\notag\\
	&\phantom{m}+m_\Lambda(m_{\Xi^*}^2-m_\Sigma^2)+m_{\Xi^*}(-4m_{\Xi^*}^2+2m_\Sigma^2+m_{\Xi^*}m_\Sigma))\notag\\
	&\phantom{mmmm}-3m_{\Xi^*}^2(m_\Lambda+2m_{\Xi^*}+m_\Sigma) s~,\notag \\
	&\tilde{G}(s)  =  3m_{\Xi^*}^2s+M_K^4+(m_\Lambda+m_{\Xi^*})(m_\Sigma+m_{\Xi^*})\notag\\
	&\phantom{mmm}\times(m_{\Xi^*}(m_{\Xi^*}-2m_\Sigma)+m_\Lambda(-2m_{\Xi^*}+m_\Sigma))\notag\\
	&\phantom{mmm}-M_K^2(m_\Lambda^2+m_\Sigma^2+2m_{\Xi^*}^2-m_{\Xi^*}(m_\Lambda+m_\Sigma))~. \notag 
\end{align}
Note that the Pascalutsa prescription of the spin-3/2 particle will bring an ambiguity in the $P_{\Sigma^*}^E$ and $P_{\Xi^*}^E$ while it keeps $K_{\Sigma^*}^E$ and $K_{\Xi^*}^E$ consistent with the interaction between the decuplet and octet states listed in Eq.~\eqref{eq:lagrangian}, see Ref.~\cite{Granados:2017cib} for the details. The uncertainties on the TFFs originating from such ambiguity, however, are negligible when compared with the parameter errors. And we take he same convention with Ref.~\cite{Granados:2017cib} where the ${\cal O}(M_\pi^2,s)$ and ${\cal O}(M_K^2,s)$ terms are dropped in the $P_{\Sigma^*}^E$ and $P_{\Xi^*}^E$. Further,
\begin{align}
	P_{0,K}^M  &=  P^M_{\rm born} + P^M_{\rm NLO}-K^M_{\Xi^*,\rm low}~,\\
	P^M_{\rm born}&=\frac34 \, \int\limits_0^\pi d\theta  \sin\theta \,\sin\theta \frac{g_B-g_A}2\frac{M2}{p_{\rm c.m.}}=g_A-g_B~,\notag\\
	P^M_{\rm NLO}&=\frac34 \, \int\limits_0^\pi d\theta  \sin\theta \,\sin\theta \left(b_{10}\frac1{F_\Phi^2}\frac2{\sqrt{3}}\right)\frac{(m_\Sigma+m_\Lambda)}{2}\frac{-M2}{p_{\rm c.m.}}
	\notag\\
	&=\frac{2}{\sqrt{3}}\frac{b_{10}}{F_\Phi^2}(m_\Lambda+m_\Sigma)~.\notag
\end{align} 
\begin{align}
	K_{K}^M  &=  K^M_{N}+K^M_{\Xi} + K^M_{\Xi^*}~,\\
	K^M_{N}&=\frac34 \, \int\limits_0^\pi d\theta  \sin\theta \,\sin\theta \bigg(-\frac{M2}{p_{\rm c.m.}}\frac12\notag\\
	&\phantom{mm}\times\bigg(\frac{g_A(m_\Lambda+m_N)(m_\Sigma+m_N)}{u-m_N^2}\bigg)\bigg)~, \notag\\
	K^M_{\Xi}&=\frac34 \, \int\limits_0^\pi d\theta  \sin\theta \,\sin\theta \bigg(-\frac{M2}{p_{\rm c.m.}}\frac12\notag\\
	&\phantom{mm}\times\bigg(-\frac{g_B(m_\Lambda+m_\Xi)(m_\Sigma+m_\Xi)}{t-m_\Xi^2}\bigg)\bigg)~, \notag\\
	K^M_{\Xi^*}&=\frac34 \, \int\limits_0^\pi d\theta  \sin\theta \,\sin\theta\frac{h_A^2}{8\sqrt{3}F_\Phi^2}\notag\\
	&\phantom{mmmmm}\times\Bigg({+\frac{M2}{p_{\rm c.m.}}\frac{1}{12m_{\Xi^*}^2(m_{\Xi^*}^2-t)}\tilde{G}(s)}\Bigg)~.\notag
\end{align} 
\begin{align}
	& K^M_{\Xi^*,\rm low} = \lim_{s \to 0} \, \lim_{m_\Lambda \to m_\Sigma} \, \lim_{M_K \to 0} K^M_{\Xi^*}(s) \notag\\
	&= \frac{h_A^2}{48 \sqrt{3} F_\Phi^2} \, 
	\frac{(-m_{\Xi^*}^2+4 m_{\Xi^*} m_\Sigma - m_\Sigma^2) \, (m_{\Xi^*} + m_\Sigma)}{m_{\Sigma^*}^2 \, (m_{\Xi^*} - m_\Sigma)} \,.
	\nonumber 
\end{align}
Here, $g_A$ and $g_B$ are defined as
\begin{align}
	g_A&=\frac1{F_\Phi^2}\left(\frac{-D}{2\sqrt{3}}+\frac{-\sqrt{3}F}{2}\right)\frac{D-F}{2}~,\notag\\
	g_B&=\frac1{F_\Phi^2}\left(\frac{-D}{2\sqrt{3}}+\frac{\sqrt{3}F}{2}\right)\frac{D+F}{2}~.\notag
\end{align}
There are only three different kinds of integration over angle involved in the $K^{E/M}$. Expanding $u$ and $t$ in the center-of-mass frame, one has
\begin{align}
	t(s,\theta)&=-\frac12 Y(s)+\frac12\kappa(s)\cos\theta,\notag\\
	u(s,\theta)&=-\frac12 Y(s)-\frac12\kappa(s)\cos\theta,\notag
\end{align} 
with $Y(s)$ and $\kappa(s)$ given by Eq.~\eqref{eq:Ykappa}. Then three different integrals are expressed as
\begin{align}
	&A=\int_{0}^\pi d\theta \frac{\sin\theta\cos\theta}{t-m_{\rm exch}^2}\frac{E1}{p_{\rm c.m.}}\propto\int_{0}^\pi d\theta \frac{\cos\theta\sin\theta}{t-m_{\rm exch}^2}\frac1{p_{\rm c.m.}}\notag\\
	&\phantom{mmm}=\frac{4}{\kappa(s)^2}-\frac{2Y(s)}{\kappa(s)^2}\tilde{K}(s),\notag\\
	&B=\int_{0}^\pi d\theta \frac{\sin\theta\cos\theta}{t-m_{\rm exch}^2}\frac{E2}{p_{\rm c.m.}}\propto\int_{0}^\pi d\theta \frac{\cos^2\theta\sin\theta}{t-m_{\rm exch}^2}\notag\\
	&\phantom{mmm}=\frac{4Y(s)}{\kappa(s)^2}-\frac{2Y(s)^2}{\kappa(s)^2}\tilde{K}(s),\notag\\
	&C=\int_{0}^\pi d\theta \frac{\sin\theta\sin\theta}{t-m_{\rm exch}^2}\frac{M2}{p_{\rm c.m.}}\propto\int_{0}^\pi d\theta \frac{\sin^2\theta\sin\theta}{t-m_{\rm exch}^2}\notag\\
	&\phantom{mmm}=\frac{-2Y(s)}{\kappa(s)^2}+\frac{Y(s)^2-\kappa(s)^2}{\kappa(s)^2}\tilde{K}(s).\notag
\end{align}
For the $u$ cases, there is an extra sign in $A$. Here we dropped all irrelevant coefficients of $\theta$-dependent terms. $m_{\rm exch}$ denotes the mass of exchanged particle, while
$\tilde{K}(s)$ is defined as~\cite{Junker:2019vvy}
\begin{equation*}
	\tilde{K}(s)=\begin{cases}
		\frac1{\kappa(s)}\log\frac{Y(s)+\kappa(s)}{Y(s)-\kappa(s)}, & (m_{\Sigma}+m_\Lambda)^2\leq s,\\
		\frac2{|\kappa(s)|}(\arctan\frac{|\kappa(s)|}{Y(s)}), & s_0\leq s\le (m_{\Sigma}+m_\Lambda)^2,\\
		\frac2{|\kappa(s)|}(\arctan\frac{|\kappa(s)|}{Y(s)}+\pi), & 4M_\pi^2\leq s\le s_0,
		\end{cases}
\end{equation*}
with $s_0=m_\Sigma^2+m_\Lambda^2+2M_\pi^2-2m_{\rm exch}^2$. Finally, $M_\pi$ is replaced by $M_K$ when calculating the expressions for the $K\bar{K}$ channel.

\bibliographystyle{elsarticle-num}
\bibliography{myrefs}

\end{document}